\documentclass[a4paper,12pt,final]{article}
\usepackage{t1enc}
\usepackage[utf8]{inputenc}
\usepackage{floatflt}
\usepackage{graphicx}
\usepackage{psfrag}
\usepackage{bbm}
\usepackage{amsmath}
\usepackage{amssymb}
\usepackage{showkeys}
\usepackage{hyperref}
\usepackage{ifthen}
\usepackage{subfigure}
\usepackage{epstopdf}
\usepackage{afterpage}

\title{
Vortices and magnetic bags in Abelian models with extended scalar sectors and some of their applications
}
\author{Péter Forgács\textsuperscript{1,2}, Árpád Lukács\textsuperscript{1}\\
{\small {}\textsuperscript{1} Wigner RCP RMI, H1525 Budapest, POB 49}\\
{\small {}\textsuperscript{2} LMPT CNRS UMR7350, Universit\'e de Tours, Parc de Grandmont, 37200 Tours, France}
}

\def\d{\mathrm{d}}
\def\e{\mathrm{e}}

\def\lag{{\mathcal{L}}}

\def\kihagy#1{}

\def\pa{\partial}

\newcommand{\arxiv}[2][]{%
  \ifthenelse{\equal{#1}{}}{%
    \href{http://arxiv.org/abs/#2}{\texttt{arXiv:#2}}%
  }{%
    \href{http://arxiv.org/abs/#2}{\texttt{arXiv:#2 [#1]}}%
  }%
}

\newcommand{\be}{\begin{equation}}
\newcommand{\ee}{\end{equation}}

\begin{document}
\maketitle

\begin{abstract}

A detailed study of vortices is presented in Ginzburg-Landau (or Abelian Higgs) models with two complex scalars (order parameters) assuming a general U(1)$\times$U(1) symmetric potential. Particular emphasis is given to the case, when only one of the scalars obtains a vacuum expectation value (VEV).  It is found that for a significantly large domain in parameter space
vortices with a scalar field condensate in their core (condensate core, CC) coexist with Abrikosov-Nielsen-Olesen (ANO) vortices. Importantly CC vortices are stable and have lower energy than the ANO ones. Magnetic bags or giant vortices of the order of 1000 flux quanta are favoured to form for the range of parameters ("strong couplings") appearing for the superconducting state of liquid metallic hydrogen (LMH). Furthermore, it is argued that finite energy/unit length 1VEV vortices are smoothly connected to fractional flux 2VEV ones.
Stable, finite energy CC-type vortices are also exhibited in the case when one of the scalar fields is neutral.
\end{abstract}

In a considerable number of physical theories describing rather different situations, vortices play often an essential r\^ole to understand key phenomena.
In gauge field theories spontaneously broken by scalar fields, the vortex of reference is undoubtedly the celebrated Abrikosov-Nielsen-Olesen (ANO) one \cite{ANO} associated to the breaking of a U(1) gauge symmetry by a complex scalar doublet. ANO vortices correspond to the planar cross-sections of static, straight, magnetic flux-tubes, with an SO(2) cylindrical symmetry.
Their magnetic flux is quantized as $\Phi=n\Phi_0$, where $\Phi_0$ is an elementary flux unit and $n$ is an integer. The integer $n$ can be identified with a topological invariant, the winding number of the complex scalar, which is also responsible for their remarkable stability. Rotationally symmetric ANO vortex solutions form families labelled by $n$ and by the mass ratio $\beta=m_s/m_v$, where $m_s$, resp.\ $m_v$ denote the mass of the scalar resp.\ of the vector field.

The ubiquity of vortices in different branches of physics, ranging from condensed matter systems, such as superfluids, superconductors \cite{Pismen, SBP, FS} to cosmic strings in high energy physics \cite{VS, kibble} greatly contributes to their importance. By now models of superconductors with {\sl several order parameters} (scalar doublets) have become the subject of intense theoretical and experimental studies \cite{ZD, Moshchalkov}. Under extremely high pressure liquid metallic hydrogen (LMH) is expected to undergo a phase transition to a superconducting state, where two types of Cooper pairs are formed, one from electrons and another one from protons \cite{Ashcroft68, Ashcroft91, Ashcroft98, Ashcroft99, Ashcroft2000, Ashcroft2005}. For experimental data on the existence of liquid metallic hydrogen see Refs.\ \cite{ET, hardpressed}, for numerical simulations, Refs.\ \cite{SSBS, simul}. Multi-component order parameters have also been considered in the context of Bose-Einstein condensates (BECs) of trapped
atoms \cite{Kasamatsu, Mason, IvashinPoluektov, KasamatsuEtoNitta, Catelani} and even for modelling of the interiors of neutron stars \cite{Jones}.

In multi-component Ginzburg-Landau (MCGL) models a number of vortex solutions differing from the ANO ones have been found. Considering two component (TCGL) theories is already sufficiently interesting for many applications in condensed matter systems, and we shall also restrict our attention to such systems in the present paper. In two-band superconductors, fractional flux vortices have been found \cite{BabaevF, BS, BCSt15, BJS, Zou, BSAdet}, with the quite remarkable property, that inter-vortex forces change their character from attractive to repulsive as the separation decreases. This phenomenon is related to type 1.5 superconductivity \cite{Moshchalkov, ZD, BS1, BS2, CBS}. Non-monotonous and non-pairwise forces also lead to the formation of vortex patterns \cite{CGB, GB}.
The purely scalar version of the theory, the two-component Gross--Pitaevskii equation,
has been applied to atomic BECs in Refs.\ \cite{Kasamatsu, Mason, IvashinPoluektov, KasamatsuEtoNitta, Catelani}. The case with one field being non-zero at the minimum of the potential was addressed for BECs in Ref.\ \cite{Catelani}, and for a TCGL with an additional $\mathbb{Z}_2$ symmetry in Ref.\ \cite{GB}. A similar, multi-component theory has also been applied to the physics of neutron star interiors in Ref.\ \cite{Jones}, where the different fields correspond to condensates of different particle species.

It has already proved fruitful to study ``universal'' properties of vortices which may turn out to be important in rather different physical settings, bringing to light analogies between condensed matter and high-energy physics.
Scalar fields with several components also appear in the standard model (SM) of particle physics, as well as in grand unified theories (GUTs), resulting in a rich catalogue of vortices \cite{VS, Witten, Peter, Peter2, vac-ach, hin1, hin2, semilocal}.
The important problem of existence and stability of vortices in the standard electroweak model, and its $\theta_W\to \pi/2$ limit, the semi-local model, have been considered in Refs.\ \cite{vac-ach, hin1, hin2, semilocal, JPV1, JPV2, Perkins, GHelectroweak}. It has been found, that ANO strings can be embedded in the SM, however, for realistic parameter values, these strings are unstable. Vortex solutions different from the embedded ones have been constructed in Refs.\ \cite{hin1, hin2, FRV1, FRV2, GVelectroweak}. It has been shown that in the semi-local model for $\beta >1$, the instability of the embedded ANO vortex corresponds to its bifurcation with a one-parameter family of solutions, which are, however, still unstable \cite{twistedinstab1, FL, twistedinstab2, GVinstab}. The existence of a lowest energy limit of this family of solutions with a less symmetric potential has been demonstrated in Ref.\ \cite{Erice}, and a short report on their stability properties can be found in Ref.\ \cite{FL2}.

The aim of the present paper is to present a detailed investigation of vortices in a class of Abelian gauge models with an extended scalar sector.
In the literature such models are referred to either as multi-component Ginzburg-Landau or extended Abelian Higgs (EAH) theories. As mentioned we shall concentrate on two-component GL theories, which already exhibit a number of interesting vortices and some interesting physics related to them.
Moreover we shall focus on the case where in the minimum of the potential one of the fields assumes a non-zero value, which has not been explored in detail up to now. We consider the most general scalar potential with a U(1)$\times$U(1) symmetry.
We expect the presented vortex solutions to be of interest both to condensed matter and to high energy physics.
The 1VEV vortices are of importance when the system is in between two different symmetry breaking transitions.
The vortices we investigate are also of relevance in a high energy physics setting when additional scalars without a VEV couple to the fields of the SM. This is the case in some models of dark matter known under the name of portal models \cite{SilveiraZee, PW}, when additional scalars are coupled to the Higgs-sector of the SM. The effect of an additional scalar on the stability of semi-local and electroweak strings has been considered in Ref.\ \cite{VachaspatiWatkins}, that of a dilaton in Ref.\ \cite{PerivolaropoulosPlatis}. The
opposite case, dark sector strings with coupling to standard matter were the subject of Refs.\ \cite{HartmannArbabzadah, BrihayeHartmann}.

Some of our main results can be summarized as follows. In purely scalar TCGL theories for a certain parameter range of the scalar potential, global vortices for $n>1$ exhibit stability (at the linear level) in sharp contrast to the known splitting instability of the single component case. This implies that the character of vortex-vortex interaction in TCGL theories changes from repulsive to attractive as the separation between the vortices decreases.

In TCGL theories with a gauge field in the 1VEV case we have investigated in detail vortex solutions and their linear stability.
The case when both scalars are charged (i.e., coupled to the gauge field) is particularly relevant for LMH and two-band superconductors. We have found that the genuinely two-component vortices with $n>1$ are stable for such parameter values when the embedded ANO one exhibits the splitting instability. Moreover their energy is always smaller than that of the embedded ANO solution.

In the case when ratio of the effective masses of the two kinds of Cooper pairs, $M=m_2/m_1$ is large (e.g., in LMH), the vortex with the smallest energy/flux ratio has a remarkably large number of flux quanta. This number of flux quanta is determined by the competition of two phenomena: (i) the condensate in the vortex core lowers the potential energy (shifting the behaviour of the system towards type I), and (ii) with the growth of the condensate in the core,
the interaction energy between the condensate and the magnetic field becomes larger (analogously to type II behaviour). The resulting ``giant'' vortices or magnetic bags are a manifestation of neither type I nor type II superconductivity.

Furthermore we explore the relationship of the 1VEV vortices with the fractional flux 2VEV ones of Refs.\ \cite{BabaevF, BS}.
We have shown that the 1VEV and 2VEV solutions are continuously connected as the parameters of the potential vary, even though
the energy of the 2VEV ones diverges.

There is another case of interest when only one of the (complex) scalars is charged. We also analyse 1VEV and 2VEV vortices and their
large flux limit in this case. It is worth to point out that in contrast to the case of two charged fields, the energy of 2VEV
vortices is {\sl finite}.

In addition to numerical studies, we present a simple analytic approximation for large flux vortices (magnetic bags), for both 1VEV cases.

The plan of the paper is as follows: in Sec.\ \ref{sec:U1xU1pot}, we recall the construction of the most general $U(1) \times U(1)$ invariant potential, and study its vacuum manifold. Two classes, 1VEV and 2VEV, are introduced, depending on the nature of the minimum. In Sec.\ \ref{sec:1VEVglob}, we consider vortices and their stability in the global model. In Sec.\ \ref{sec:theory}, two-component EAH models are introduced, and twisted vortices are studied. In Sec.\ \ref{sec:zerotwist} stable condensate core vortices are studied numerically. Main stability results are contained in Sec.\ \ref{sec:linpert}. Large flux vortices are studied in detail in Sec.\  \ref{sec:limits}. In Sec.\ \ref{sec:2VEV} we present our results in the 2VEV case, with emphasis on the large $M$ limit. Conclusions are presented in Sec.\ \ref{sec:conclusions}. Details of the calculations are relegated to Appendix \ref{app:pertdetails}, and some numerical data to Appendix \ref{sec:numdata}.

\section{\texorpdfstring{$U(1)\times U(1)$}{U(1) x U(1)} invariant potentials and their vacua}\label{sec:U1xU1pot}

Considering two complex scalar fields the most general $U(1)\times U(1)$ symmetric self-interaction potential has already been given in Ref.\ \cite{Witten}:
\begin{equation}\label{eq:pot}
  V = \frac{\beta_1}{2} (|\phi_1|^2-1)^2 + \frac{\beta_2}{2} |\phi_2|^4 + \beta' |\phi_1|^2|\phi_2|^2 - \alpha |\phi_2|^2\,,
\end{equation}
containing four real parameters, $\beta_1$, $\beta_2$, $\beta'$ and $\alpha$.
In the rest of our paper we shall consider theories where the potential, $V$, is given by \eqref{eq:pot}, moreover we require that $V>0$ for $|\phi_1|^2\,,|\phi_2|^2\to\infty$, resulting in the following restriction of the parameters: $\beta_1>0$, $\beta_2>0$ and $\beta'>-\sqrt{\beta_1\beta_2}$.

The following two types of minima of the potential \eqref{eq:pot} shall be considered: either a state when only a single scalar field has a VEV, referred to as 1VEV case, and the other one is a 2VEV case where both fields obtain a VEV.
The conditions for a 2VEV state is
\begin{equation}\label{eq:2VEVcond}
\alpha > \beta'\,,\quad \beta_1 \beta_2 > \alpha \beta'\,,
\end{equation}
from which $\beta_1 \beta_2 > {\beta'}^2$ follows. In this 2VEV case, the two vacuum expectation values of the scalar fields, $\eta_1\,,\eta_2$ satisfy
\begin{equation}\label{eq:2VEV}
  \eta_1^2 = \frac{\beta_1\beta_2-\alpha \beta'}{\beta_1\beta_2-(\beta')^2}\,,\quad
  \eta_2^2 = \frac{\beta_1(\alpha- \beta')}{\beta_1\beta_2-(\beta')^2}\,,
\end{equation}
and the previous conditions guarantee that $\eta_1^2\,,\eta_2^2>0$.

If at least one of the conditions in Eq.\ (\ref{eq:2VEVcond}) fails to hold, the system is in a 1VEV state, and the component having the non-zero VEV is as follows:
\begin{equation}\label{eq:VEVcond1}
\begin{tabular}{c| c| c}
          & $\beta_1 \beta_2 > {\beta'}^2$   & $\beta_1 \beta_2 < {\beta'}^2$\\
\hline
upper     & $\beta' > \alpha$              & $\sqrt{\beta_1 \beta_2} > \alpha$\\
\hline
lower     & $\beta' < \alpha$              & $\sqrt{\beta_1 \beta_2} < \alpha$\\
\end{tabular}
\end{equation}
The classification in Eqs.\ (\ref{eq:2VEVcond}), (\ref{eq:VEVcond1}) is crucial. If $\beta_1\beta_2 > {\beta'}^2$,  then at $\alpha = \beta'$, there is
a boundary between the upper component 1VEV and the 2VEV cases. If, on the contrary,  $\beta_1\beta_2 < {\beta'}^2$ then at $\alpha = \sqrt{\beta_1 \beta_2}$,
there is a boundary between upper component and lower component 1VEV cases, there the lower component obtains a VEV $\eta_2 = \sqrt{\alpha/\beta_2}$.

In deriving Eq.\ (\ref{eq:VEVcond1}), we have used that in order that the global minimum of the potential be at the field values $(1,0)$,
\begin{equation}\label{eq:cond1stab1}
V(\phi_1=0,\phi_2=\eta_2)=\frac{\beta_1}{2}-\frac{\alpha^2}{2\beta_2} > 0
\end{equation}
has to hold with $\eta_2^2=\alpha/\beta_2$.
Condition (\ref{eq:cond1stab1}) corresponds to that out of the two possible
local minima $(1,0)$ and $(0,\eta_2)$ the first one be the global minimum.
This can be assumed without loss of generality (because otherwise the second component would be the one obtaining a VEV,
and the two components could be interchanged).

\section{Global vortices with a single VEV }\label{sec:1VEVglob}
Let us start by considering the two component scalar theory with interaction potential \eqref{eq:pot}, admitting a global
U(1)$\times$U(1) symmetry, defined by the Lagrangian
\begin{equation}
 \label{eq:globlag}
 \mathcal{L}_{\rm glob} = \partial_\mu \Phi^\dagger \partial^\mu \Phi - V(\Phi^\dagger, \Phi)\,,
\end{equation}
where $\Phi=(\phi_1,\phi_2)^T$ and $\Phi^\dagger=(\phi_1^*,\phi_2^*)$, the potential, $V$, is given by Eq.\ \eqref{eq:pot}.
We shall now consider global vortex solutions of the theory \eqref{eq:globlag} with rotational symmetry in the plane,
with the following (standard) Ansatz for the scalars
\begin{equation}
 \label{eq:globAns}
 \phi_1 = f_1(r) \e^{i n\vartheta}\,,\quad \phi_2 = f_2(r) \e^{i m \vartheta}\,,
\end{equation}
where $n$ and $m$ are integers and $(r,\vartheta)$ are the polar coordinates in the plane.
The radial equations read in this case
\begin{equation}
 \label{eq:globradeq}
 \begin{aligned}
  \frac{1}{r}(r f_1')' &= f_1 \left[ \frac{n^2}{r^2} + \beta_1 (f_1^2-1) + \beta' f_2^2 \right]\,,\\
  \frac{1}{r}(r f_2')' &= f_2 \left[ \frac{m^2}{r^2} + \beta_2 f_2^2 - \alpha + \beta' f_1^2 \right]\,,
 \end{aligned}
\end{equation}
and with the Ansatz (\ref{eq:globAns}), the energy density from the Lagrangian in Eq.\ (\ref{eq:globlag}) is
\begin{equation}
 \label{eq:globerg0}
 \mathcal{E} = (f_1')^2 + (f_2')^2 + \frac{n^2}{r^2}f_1^2 + \frac{m^2}{r^2}f_2^2  + V\,.
\end{equation}
In the following we shall focus on vortices with $m=0$, as they are expected to give solutions of ``lowest'' energy.
In the present 1VEV case embedded global ANO-type vortices, $(\phi_1,\phi_2)=(\phi_1^{(n)},0)$  automatically satisfy
Eqs.\ \eqref{eq:globradeq}. As it is known, the total energy of global vortices diverges:
\begin{equation}
 \label{eq:globerg}
 E(R) = 2\pi \int_0^R \mathcal{E} r\d r \sim E(R_{\rm core}) + 2\pi n^2 \log\left( \frac{R}{R_{\rm core}} \right)\,,
\end{equation}
where $R_{\rm core}$ is an arbitrary (core) radius, outside of which all fields can be replaced with their asymptotic form. The energy of the vortices diverges logarithmically with the sample size.

\begin{figure}
\centering
\noindent\hfil\includegraphics[scale=.5,angle=-90]{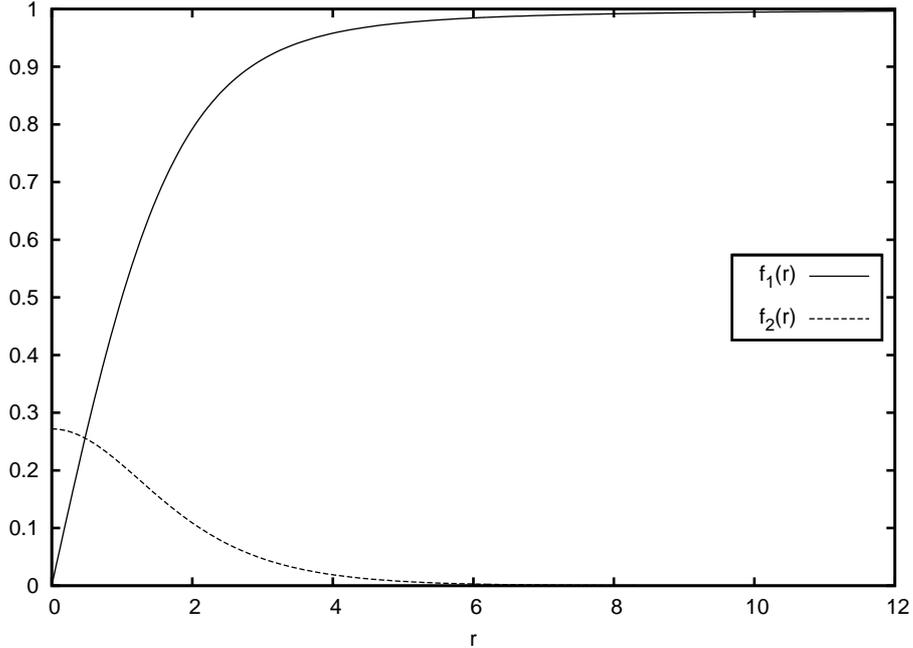}
\caption{A global CC vortex for $\beta_1=1$, $\beta_2=2$, $\beta'=2$ and $\alpha=1.24$.}
\label{fig:globcc}
\end{figure}

As a non-zero $\phi_2$ lowers the potential energy in the vortex core,
it is natural to expect that Eqs.\ \eqref{eq:globradeq} may also admit vortex solutions with a non-trivial $\phi_2$.
A simple method to search for such non-trivially two-component vortices is to look for the instability of the embedded one.
This can be done by linearising Eqs.\ \eqref{eq:globradeq} around an embedded vortex in the small parameter $\epsilon = f_2(0)$ as follows:
\begin{equation}
 \label{eq:bifur-global-expan}
 \begin{aligned}
    \alpha &=\alpha_{\rm b} + \epsilon^2 \alpha_2 + \dots\,,\\
    f_1    &= f_1^{(0)} + \epsilon^2 f_1^{(2)} + \dots\,,\\
    f_2    &= \epsilon f_2^{(1)} + \dots\,,
 \end{aligned}
\end{equation}
and $f_2^{(1)}$ satisfies the following linear Schr\"odinger-type equation, with a ''potential'' determined by the embedded vortex, $ f_1^{(0)}$:
\begin{equation}
 \label{eq:bifur-global}
 \frac{1}{r}(r f_2^{(1)}{}')' - \beta' f_1^{(0)2}f_2^{(1)} = \alpha_{\rm b}f_2^{(1)}\,.
\end{equation}
In order that one obtain a linearized vortex solution, one has to impose $f_2^{(1)}\to0$ for $r\to\infty$.
Then Eq.\ \eqref{eq:bifur-global} can be interpreted as an eigenvalue problem for the ``energy'' $\alpha_{\rm b}$ as a function of the parameter $\beta'$.
For the parameter range $\alpha_{\rm b} < \alpha < \beta'$, embedded global vortices are unstable, and they bifurcate with a new family of solutions with a nontrivial $f_2$ as $\alpha \to\alpha_{\rm b}$, to which we shall refer to as condensate core (CC) vortices. A numerical solution for a global CC vortex is depicted on Fig.\ \ref{fig:globcc}. Although the energy of a CC vortex is also divergent logarithmically, just as for an embedded one, for a fixed sample size $R$, the energy difference between the two types can be computed.
It turns out that CC vortices have lower energy than the embedded ones, and in some cases the energy difference is remarkably
large (Table \ref{tab:globcc}).

 Remarkably good approximate solutions of Eq.\ \eqref{eq:bifur-global} are known for both large and small values of $\beta'$. In the case when $\beta'\gg\beta_1$, the lowest eigenfunction is concentrated close to the origin, therefore a good approximation of the potential term is only needed there.  As noted in Ref.\ \cite{StringsSprings} the potential to leading order is harmonic yielding
 a qualitatively good approximation of Eq.\ \eqref{eq:bifur-global}.
 The harmonic approximation can be substantially improved by taking into account the Taylor expansion of $f_1^2$ up to the $r^6$ order via perturbation theory \cite{Catelani}, yielding
\begin{equation}
 \label{eq:alphablargebetap}
 \alpha'_{\rm b} \approx 2\sqrt{\beta'} - \frac{1}{2} + \frac{5+16 c_0^2}{32 c_0} {\beta'}^{-1/4}\,,\quad \left(\beta' > \frac{\beta_1}{3}\right)\,,
\end{equation}
where $c_0=f_1'(0)$. For small $\beta'$, a matching procedure at the boundary of the vortex core yields the eigenvalue \cite{Catelani}
\begin{equation}
 \label{eq:alphabsmallbetap}
 \alpha'_{\rm b} \approx \beta' - \frac{4}{\beta'} \e^{-2\gamma_{\rm E}} c_0^2 \exp\left[ {-\frac{2}{\sqrt{\beta'}} \arctan\frac{2 c_0^2}{\sqrt{\beta'}}}\right]\,,\quad \left(\beta' < \frac{\beta_1}{3}\right)
\end{equation}
where $\gamma_{\rm E}\approx 0.5772$ is the Euler-Mascheroni constant. A similar result has been obtained in Ref.\ \cite{StringsSprings} for the case of gauged vortices, based on approximate eigenvalues of shallow potentials in 2D \cite{Landau3}.

In Ref.\ \cite{Goodband}, it has been demonstrated numerically, that global vortices for $n>1$ are unstable against splitting into vortices of lower winding. This is in agreement with the known {\sl repulsive} interaction between global vortices at large separations
\cite{Pismen}.
For CC vortices, the leading order asymptotic behaviour is unchanged, therefore at large separation there should be a repulsive interaction between them. On the other hand, for CC vortices close to each other, the nonzero second component also contributes to the inter-vortex force. We have performed a stability analysis with the methods of Ref.\ \cite{Goodband}.
We have found that for parameter values of $\alpha$ away from the bifurcation $\alpha\gg\alpha_{\rm b}$, $n=2,3$ CC vortices are stable at the linear level.
In the case of the embedded vortex, for $n=2$, there is an energy lowering perturbation (an eigenfunction of the perturbation operator with a negative eigenvalue) in the partial wave channel $\ell=2$, and for $n=3$ in $\ell=2,3,4$. For the CC vortex, sufficiently far from the bifurcation, these eigenvalues become positive. We denote by $\alpha_{\rm s}$ the value of $\alpha$ where all of the eigenvalues become positive. See Tab.\ \ref{tab:globstab} for numerical data.
See also Sec.\ \ref{sec:linpert} and Appendix \ref{app:pertdetails} for details of the method.

It is remarkable, that the character of the inter-vortex force changes in the two-component theory, from attractive at small separations to repulsive at large ones. This is analogous to the behaviour of vortices in type 1.5 superconductors.

\begin{table}
  \centering
  \begin{tabular}{|c|c|c||c|}
  \hline
    $\beta_2$ & $\beta'$ & $\alpha$ & $E_{\rm e} - E_{\rm cc}$ \\
    \hline\hline
      1       & 1       & 0.94      & 2.161 \\
      2       & 2       & 1.24      & $8.795 \times 10^{-3}$ \\
      50      & 10      & 6.2       & 0.8864 \\
    \hline
  \end{tabular}
 \caption{Energy difference between embedded and condensate core global vortices, $\beta_1=1$.}
 \label{tab:globcc}
\end{table}

\begin{table}
\centering
\begin{tabular}{|c||c|c|}
 \hline
 $n$ & $\alpha_{\rm b}$ & $\alpha_{\rm s}$ \\
 \hline\hline
 1   & 1.2052           & ---    \\
 2   & 0.60143          & 1.3208 \\
 3   & 0.34771          & 1.3454 \\
 \hline
\end{tabular}
 \caption{Stabilisation of global CC vortices: for $\alpha_{\rm s} < \alpha < \beta'$, no negative eigenvalues were found. Here $\beta_1=1$, $\beta_2=\beta'=2$.}
 \label{tab:globstab}
\end{table}

\section{Twisted vortices in Extended Abelian Scalar models}\label{sec:theory}
The Lagrangian of the two-component EAH model (the relativistic version of the TCGL) is
\begin{equation}
  \label{eq:Lag}
  \lag_{\rm lok} = \frac{1}{e^2}\left\{-\frac{1}{4} F_{\mu\nu}F^{\mu\nu} + (D_\mu\Phi)^\dagger (D^\mu\Phi) - V(\Phi,\Phi^\dagger) \right\}\,,
\end{equation}
where $D_\mu\phi_a = \partial_\mu - i e_a A_\mu$ is the standard gauge covariant derivative of the scalars, where for later use we assume general couplings, $(e_1\,,e_2)$, of $\Phi=(\phi_1, \phi_2)^T$ to the U(1) gauge field, and $V$ is defined by Eq.\ (\ref{eq:pot}).

The $U(1)$ gauge symmetry acts on the fields as
$\Phi \to \exp(i\chi)\Phi$, $A_\mu \to A_\mu + \partial_\mu \chi$, where $\chi=\chi(x)$ is the gauge function. The other $U(1)$ symmetry is global, and it acts on the fields as $\phi_1 \to \exp(-i\alpha)\phi_1$, $\phi_2 \to \exp(i\alpha)\phi_2$, where $\alpha$ is a constant.

The field equations obtained from the Lagrangian (\ref{eq:Lag}) read
\begin{equation}
  \label{eq:EOM}
  \begin{aligned}
    \pa^\rho F_{\rho\mu}& =i\sum_a e_a\{(D_\mu\phi_a)^*\phi_a -\phi_a^* D_\mu\phi_a\}\,, \\
    D_\rho D^\rho\Phi & =  -\pa V(\Phi^\dagger,\Phi)/\pa \Phi^\dagger \,.\\
   \end{aligned}
\end{equation}

The theory defined in Eq.\ (\ref{eq:Lag}) is a member of the family of semilocal models, i.e., gauge theories with additional
global symmetries \cite{semilocal}. A thoroughly studied case is the $SU(2)_{\text{global}}\times U(1)_{\text{local}}$ semilocal model \cite{semilocal}, corresponding to the limit $\theta_W\to \pi/2$ of the standard electro-weak model corresponding to the parameter choice
$\beta_1=\beta_1=\beta'=\alpha$.

Importantly, in the 1VEV case, solutions of the ordinary one-component Abelian Higgs model can be embedded in the theory, as $\phi_1=\phi_{AH}$,
$\phi_2=0$ and $A_\mu=A_{\mu,AH}$, where $\phi_{AH}, A_{\mu,AH}$ is a solution of the one-component model with $\beta=\beta_1$. In this way, we can consider embedded ANO vortices in the 1VEV two-component theory.

The conserved current corresponding to the global $U(1)$ symmetry of the theory (\ref{eq:Lag}) is given by
\begin{equation}
  \label{eq:jmu}
  j_\mu^{3} = -i ( \phi_1^* D_\mu \phi_1 - \phi_2^* D_\mu \phi_2  - \phi_1 (D_\mu\phi_1)^* + \phi_2 (D_\mu\phi_2)^*)\,,
\end{equation}
which agrees with the third isospin component of the global $SU(2)$ current of the semilocal theory \cite{FRV1, FRV2}.

The general stationary, cylindrically symmetric Ansatz introduces $z$-dependent phases for the scalars, and a suitably reduced Ansatz in the radial gauge can be written as
\begin{equation}
  \label{eq:FRVAns}
\begin{aligned}
  \phi_1(r,\vartheta,z)\, &= f_1(r) e^{i n\vartheta}\,, \\
  \phi_2(r,\vartheta,z)\,  &= f_2(r) e^{i m\vartheta}e^{i\omega z}\,, \\
\end{aligned} \quad \begin{aligned}
  A_\vartheta(r,\vartheta,z) &= n a(r)\,,\\
  A_3(r,\vartheta,z) &= \omega a_3(r)\,,
\end{aligned}
\end{equation}
with $A_0=A_r=0$ and $\omega$ is real, it shall be referred to as the twist parameter.
 The Ansatz (\ref{eq:FRVAns}) describes cylindrically symmetric fields in the sense, that a translation along the $z$ direction can be compensated by the application of internal symmetries \cite{FM, FRV1, FRV2}. All twisted solutions, where the spacetime dependence of the relative phase is timelike, can be brought to the form of Eq.\ \eqref{eq:FRVAns} by a Lorentz boost.

With the Ansatz, Eq.\ (\ref{eq:FRVAns}), the field equations, Eq.\ (\ref{eq:EOM}) become
\begin{equation}
  \label{eq:FRVprofE}
  \begin{aligned}
    \frac{1}{r}(r a_3')' &= 2 a_3 (e_1^2 f_1^2 + e_2^2 f_2^2) - 2 e_2 f_2^2,\\
    r\left(\frac{a'}{r}\right)' &= 2 f_1^2 e_1 ( e_1 a - 1 ) + 2 f_2^2 e_2 (e_2 a - m / n ) \,,\\
    \frac{1}{r}(r f_1')' &= f_1\left[ \frac{(1 - e_1 a)^2 n^2}{r^2} + e_1^2 \omega^2 a_3^2 + \beta_1 ( f_1^2 - 1 ) + \beta' f_2^2\right]\,,\\
    \frac{1}{r}(r f_2')' &= f_2\left[ \frac{(e_2 n a-m)^2}{r^2} + \omega^2 (1-e_2 a_3)^2 + \beta_2 f_2^2 -\alpha + \beta' f_1^2 \right]\,.
  \end{aligned}
\end{equation}
The boundary conditions for regular, 1VEV solutions of Eqs.\ (\ref{eq:FRVprofE}) imply that $f_1(r=0)=0$, and for $m=0$ $f_2(r=0)=$const.\, while for $r\to\infty$ we impose that $f_1,a\to 1$ and $f_2,a_3\to 0$.
In the 2VEV case, $f_{1,2} \to \eta_{1,2}$, where $\phi=(\eta_1,\eta_2)$ is a minimum of $V$.
In this latter case, twisted vortex solution would have infinite energy (proportional to the sample volume), since the
one cannot satisfy  for $(r\to\infty)$ both $D_3\phi_1\to0$ and $D_3\phi_2\to0$ simultaneously.
We start with the description of finite energy twisted vortex solutions of Eqs.\ \eqref{eq:FRVprofE}, therefore we impose  $f_2\to 0$ for $(r\to\infty)$.

The energy density for the Ansatz \eqref{eq:FRVAns} is found to be
\begin{equation}
  \label{eq:Edens}
  \begin{aligned}
  \mathcal{E} =& \frac{1}{2}\left[ \frac{n^2 (a')^2}{r^2} + \omega^2 (a_3')^2\right]
  + (f_1')^2 + (f_2')^2 \\
  &+ \frac{n^2(1 - e_1 a)^2}{r^2} f_1^2 + \frac{(e_2 na - m)^2}{r^2} f_2^2 + \omega^2(e_1^2 a_3^2 f_1^2 + (1-e_2 a_3)^2 f_2^2) + V(f_1,f_2)\,.
  \end{aligned}
\end{equation}
with $V(f_1,f_2)=\beta_1 (f_1^2-1)^2/2 + \beta_2f_2^4/2 + \beta' f_1^2f_2^2 - \alpha f_2^2$.
The total energy (per unit length), $E$, is given as the integral over the plane of $\mathcal{E}$,
\begin{equation}
 \label{eq:toterg}
 E = 2\pi\int_0^\infty r\d r \mathcal{E}\,.
\end{equation}
$E$ is a monotonously increasing function of the parameters $\beta_1$, $\beta_2$, $\beta'$, and of the twist, $\omega$, while it is a monotonously decreasing function of $\alpha$. This follows from the fact that if $\Phi, A_\mu$ is a static solution of the field equations, then, e.g.,
\begin{equation}
 \label{eq:Ederivomega}
 \frac{\partial E}{\partial \omega^2} = 2\pi \int r \d r \left[ \frac{1}{2}(a_3')^2 + (e_1^2 a_3^2 f_1^2 + (1-e_2 a_3)^2 f_2^2) \right] > 0\,.
\end{equation}
Some curves depicting the total energy as a function of the twist for some solution families are shown on Figure \ref{fig:Eomega}.

Plugging the Ansatz (\ref{eq:FRVAns}) into (\ref{eq:jmu}), the relevant current component is
\begin{equation}
  \label{eq:j33}
  j_3^{3} = 2 \omega a_3(e_1 f_1^2 - e_2 f_2^2) + 2 \omega f_2^2\,.
\end{equation}
The global current, $\mathcal{I}(\omega)$, is depicted on Fig.\ \ref{fig:j3omega}, where
the $SU(2)$ symmetric case, $\beta_{1,2}=\beta'=\alpha=2$ is compared to a less symmetric one, for $\beta_{1,2}=\alpha=2$, $\beta'=2.1$.
In the $SU(2)$ symmetric case, $\mathcal{I}(\omega)$ diverges for $\omega\to 0$ \cite{FRV1, FRV2}, and there is no
finite energy solution corresponding to $\omega=0$. As we shall demonstrate in the general, nonsymmetric case, finite energy vortex solutions exist in the $\omega\to0$ limit.

The numerical solutions of Eqs.\ (\ref{eq:FRVprofE}) have been calculated using the shooting method with a fitting point \cite{NR}, which is also used for the solution of the linearised equations for the stability analysis. For higher winding number vortices, we also use a minimisation of the energy functional (\ref{eq:Edens}) directly, in a finite difference discretisation.

\subsection{Bifurcation with embedded ANO strings}\label{ssec:bifurcation}
It is by now well known that embedded ANO vortices are unstable to small perturbations of the $f_2$ variable \cite{hin1, hin2},
and that this instability corresponds to the aforementioned bifurcation \cite{FRV1, FRV2}. Close to the bifurcation, a systematic expansion of
the solution in a bifurcation parameter $\epsilon$ has been carried out in Ref.\ \cite{FL} in the $SU(2)$ symmetric case.
The analysis of Ref.\ \cite{FL} can be repeated in the present case with minimal modifications.
The systematic expansion of a twisted vortex near the bifurcation point
can be then written as :
\begin{equation}
  \label{eq:bifureps}
  \begin{aligned}
    f_1 &= f_1^{(0)} +\epsilon^2 f_1^{(2)} + \ldots \\
    f_2 &= \epsilon f_2^{(1)}+ \epsilon^2 f_2^{(2)}+\ldots \\
\end{aligned}\quad \begin{aligned}
    a   &=  a^{(0)}  + \epsilon^2 a^{(2)} + \ldots \\
    a_3 &=  \epsilon^2 a_3^{(2)} + \ldots \\
    \omega &= \omega_{\rm b} + \epsilon^2 \omega_2\,\,\, + \ldots
  \end{aligned}
\end{equation}
where $a^{(0)}\,,f_1^{(0)}$ denotes the ANO vortex, whose equations can be read off
from equations (\ref{eq:FRVprofE}) by putting $f_2=a_3=0$.\
For details, and the Taylor expanded equations, see \cite{FL}.

The leading order equation is
\begin{equation}\label{eq:bifur-f2}
(D_2^{(0)}+ \omega_{\rm b}^2)f_2^{(1)}
  := -\frac{1}{r}\left(r {f_2^{(1)}}'\right)' + \left[ \frac{(e_2 na^{(0)}-m)^2}{r^2}
 + \omega_{\rm b}^2-\alpha +\beta'(f_1^{(0)})^2\right] f_2^{(1)}=0\,.
\end{equation}
The expansion coefficients $\omega_i$ are dictated by the conditions for the cancellation of resonance terms. The procedure
yields $\omega_1=0$, thus
\begin{equation}
\epsilon = \sqrt{\frac{1}{\omega_2}(\omega-\omega_{\rm b})} + \dots\,.
\end{equation}

\paragraph{Energy difference} 
Twisted vortices have lower energies than embedded ANO ones (see Subsection \ref{ssec:1VEV11} for numerical values), and in some cases, this energy difference is remarkably large.
The explanation is, that in the core of an embedded vortex, there is false vacuum, which, in the case of a twisted vortex, is filled with the second condensate, reducing the potential enegy. This also has costs in the form of derivative and interaction terms. In those cases, where $f_2 \ll 1$ [e.g., close the the bifurcation ($\omega \approx \omega_{\rm b}$)], this energy difference can be calculated approximately,
with the help of the bifurcation equation, Eq.\ (\ref{eq:bifur-f2}) \cite{Catelani}. Neglecting the term quartic in $f_2$, and the backreaction of $f_2$ on $f_1$ in the energy density (\ref{eq:Edens}),
and performing a partial integration yields
\begin{equation}
 \label{eq:echange}
 E - E_{ANO} \approx 2\pi\int r \d r f_2 \left\{ -f_2'' - \frac{1}{r}f_2' + f_2 \left[ \frac{(e_2na-m)^2}{r^2} -\alpha + \beta' f_1^2 \right]\right\} = -\omega_{\rm b}^2 2\pi\int r\d r f_2^2\,.
\end{equation}

\subsection{Numerical solutions}\label{ssec:1VEV11}
Let us first consider the case of $e_1=e_2=1$. The $SU(2)$ symmetric case has been considered in Refs.\ \cite{vac-ach, hin1, hin2, semilocal}.

The range of solutions can be found by solving the bifurcation equation, Eq.\ (\ref{eq:bifur-f2}). In the $SU(2)$ symmetric
case, for $\beta_1 > 1$, an instability is found. In these cases, twisted vortices exists for $ 0 < \omega < \omega_{\rm b}$. For some parameter values,
$\omega_{\rm b}$ is shown in Table \ref{tab:bif_e21}.

\begin{figure}
\centering
\subfigure[{}]{
\includegraphics[scale=.32,angle=-90]{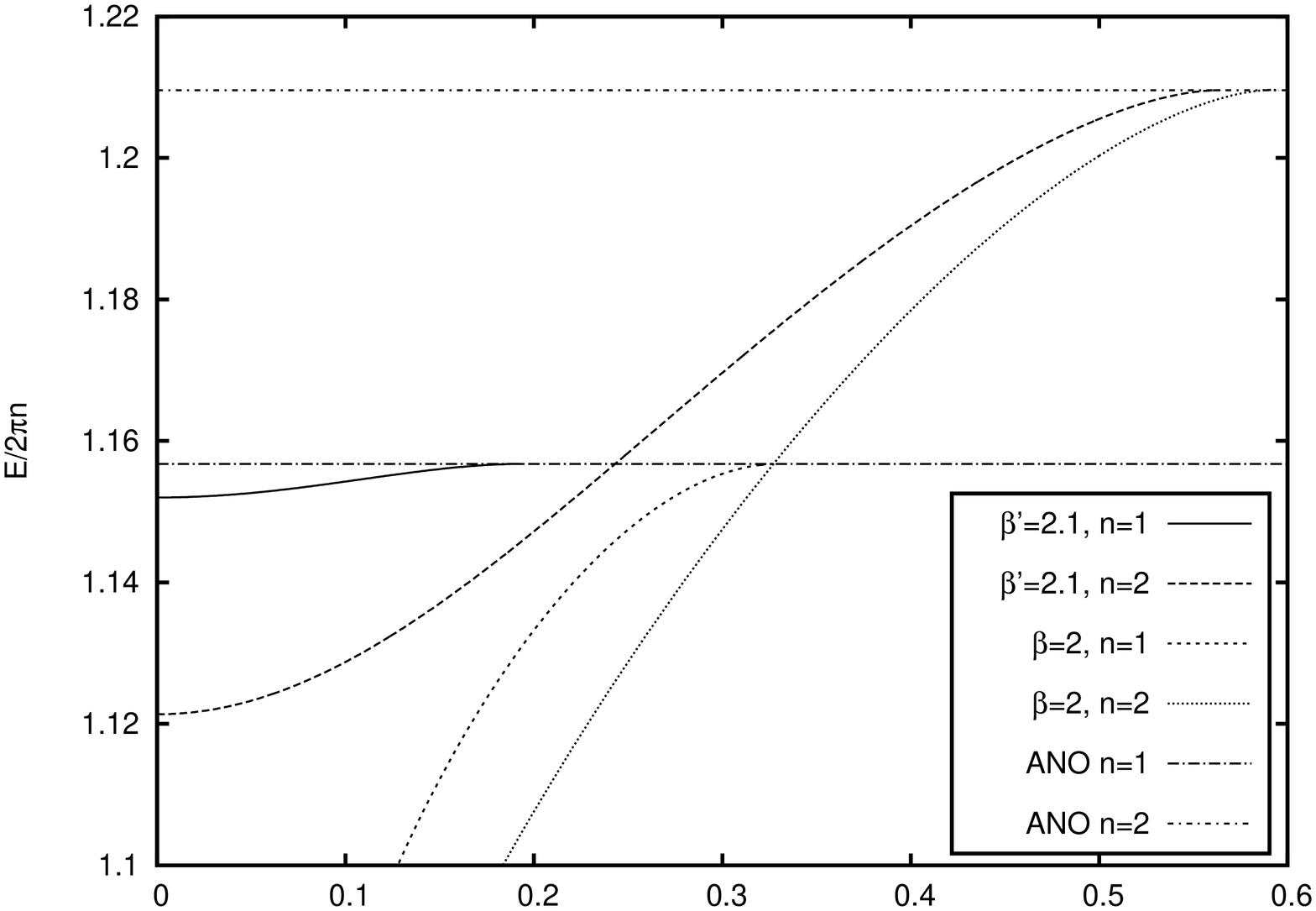}
\label{fig:Eomega}
}
\subfigure[{}]{
\includegraphics[scale=.32,angle=-90]{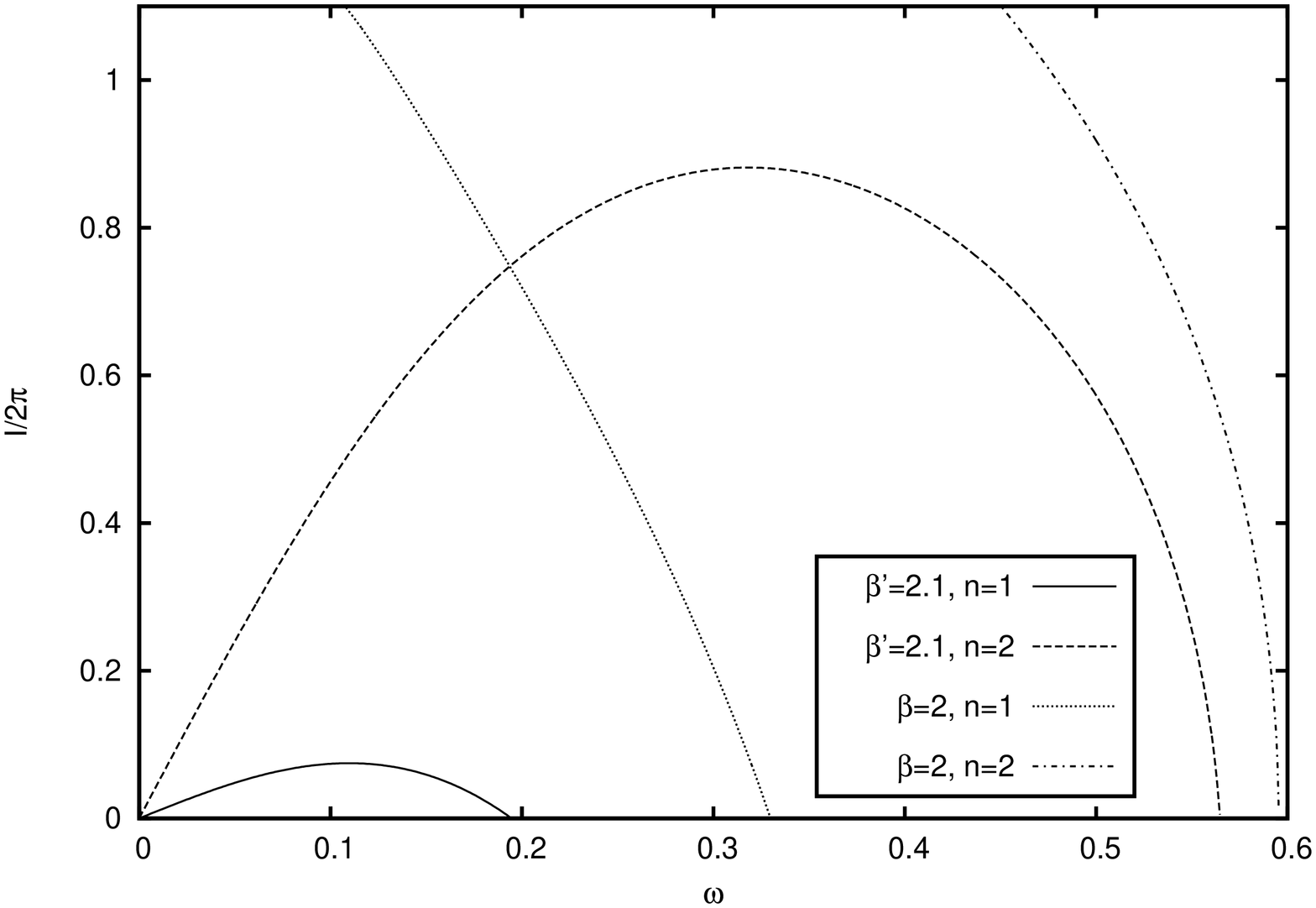}
\label{fig:j3omega}
}
\caption{The energy $E$ and the current, $\mathcal{I}$ as a function of the twist $\omega$}
\label{fig:Ej3omega}
\end{figure}

In Fig.\ \ref{fig:Eomega}, the dependence of the vortex energy on the twist is displayed, and
the dependence of the global current $\mathcal{I}$ on the twist $\omega$ is depicted in figure \ref{fig:j3omega}. See also the $SU(2)$ symmetric case in Ref.\ \cite{FRV1, FRV2}.

Numerically we have found that twisted string solutions exist for $0<\omega<\omega_{\rm b}$, where the upper limit is a function of the parameters $\beta_1$, $\beta_2$, $\beta'$ and $\alpha$ of the potential and the flux $n$ of the vortex, similarly to the $SU(2)$ symmetric case \cite{FRV1, FRV2} (we have assumed $m=0$).

In the case of one charged and one neutral fields, $e_2=0$, as seen from Eqs.\ (\ref{eq:FRVprofE}), $a_3=0$.
In both the field equations and
the energy, Eq.\ (\ref{eq:Edens}), the same profile functions and energy is obtained with the replacement $\omega \to 0$
and $\alpha \to \alpha -\omega^2$, with the global current $j^3_2 = 2 \omega f_2^2$.
Therefore, for $e_2=0$ twisted vortices can be considered as trivial transformations of zero twist ones.
[A similar argument applies to the case of the global theory (Sec. \ref{sec:1VEVglob}) as well.]

\begin{table}
 \centering
 \begin{tabular}{|c|c|c||c|}
  \hline
  $\beta_1$ & $\beta'$ & $\alpha$ & $\omega_{\rm b}$ \\
  \hline\hline
  1.25      & 1.25     & 1.25     & 0.13667 \\
  2         & 2        & 2        & 0.32989 \\
  2.5       & 2.5      & 2.5      & 0.42744 \\
  1.25      & 1.255    & 1.25     & 0.12100 \\
  2         & 2.1      & 2        & 0.19453 \\
  2.5       & 2.6      & 2.5      & 0.33726 \\
  \hline
 \end{tabular}
\caption{The value of the twist at the bifurcation for $e_2=1$.}
\label{tab:bif_e21}
\end{table}

\section{Condensate core vortices}\label{sec:zerotwist}
The $\omega\to 0$ limit of twisted vortices is quite remarkable: as the energy is a monotonous function of the twist, assuming its maximum at the embedded vortices, $\omega=\omega_{\rm b}$,
the zero twist limit, i.e., condensate core, or coreless \cite{Catelani} vortices are minimum energy solutions, coexisting with embedded ANO vortices, with energies, in some cases significantly, lower.

If $\alpha < \beta'$, they are exponentially localised, $f_2 \sim F_2 r^{-1/2} \exp{-\sqrt{\beta'-\alpha}r}$, where $F_2$ is a constant, determined by the global solution of the bounday value problem [i.e., by the numerical solution of the radial equations, Eq.\ (\ref{eq:FRVprofE})].
As minimum energy solutions, they are expexted to be stable: $n=1$ ANO vortices in this theory are known to have one negative eigenvalue mode, the one corresponding to the bifurcation \cite{hin1, hin2, FRV1, FRV2}. For $n>1$, ANO vortices are also unstable for $\beta_1>1$ against decay into lower winding number ones, therefore, for CC vortices, $n>1$ requires a numerical investigation of the linearised equations.

In the case of ANO vortices, the instability of higher flux vortices for $\beta >1$ is a consequence of the repulsive interaction between unit flux ones. That the change in the stability occurs at $\beta=1$
follows from the fact that for $\beta < 1$, the scalar field has a slower radial fall-off, $\sim F r^{-1/2} \exp(-\sqrt{2\beta}r)$, than the gauge field, $\sim A r^{1/2} \exp(-\sqrt{2}r)$,
whereas for $\beta < 1$, the scalar field falls off more slowly, and the interaction is attractive. Here, for a wide range of parameters, the second scalar has the slowest radial fall-off, and thus
the interaction between two vortices can be attractive even if $\beta_1 > 1$.

The existence of zero twist vortices if $\beta'=\alpha$ is also possible. In the $SU(2)$ symmetric model, no such solutions exist for $\beta >1$, although a consistent asymptotic solution can be found. As $\omega\to 0$, vortices become diluted \cite{FRV1, FRV2}. In the $\beta=1$ case, there is a one parameter family of solutions with degenerate energy \cite{hin1, hin2}. In the non-symmetric case, we have found that if $\beta_1 \beta_2 \ne {\beta'}^2$, zero twist CC vortices still exist, with a power law asymptotic behaviour, $f_2 \sim F_2/r$, where $F_2$ is a constant. 
See also Ref.\ \cite{GB} for the $U(1)\times U(1)\times\mathbb{Z}_2$ symmetric case. In the latter case, due to the high degree of symmetry of the potential, a domain structure also exists.

If $\beta_1 \beta_2 \ne (\beta')^2$, the CC vortices, continued into the range $\alpha > \beta'$ (2VEV), become the 2VEV vortices with winding in the upper component.
If $e_2\ne 0$, these are fractional flux vortices of Refs.\ \cite{BabaevF, BS}. We shall briefly return to the 2VEV case in Sec.\ \ref{sec:2VEV}.
In the $\beta_1\beta_2 ={ \beta'}^2$, $\alpha=\beta'$ case, there seems to be no limiting solution. In these cases, as the twist
$\omega$ decreases, the profile functions reach their asymptotic values farther from the origin. This way, the string expands and its
energy density becomes more dilute. In Ref.\ \cite{FRV1, FRV2}, this behaviour has
been described with a scaling argument in the $SU(2)$ symmetric case,
which can be generalized to the $\beta_1 \beta_2 = (\beta')^2$ case without major changes.

If $\beta_1\beta_2 = \alpha^2$ (the boundary between upper and lower component 1VEV), solutions with the upper and the lower component having a non-zero VEV coexist.
For the special case of $U(1)\times U(1)\times\mathbb{Z}_2$ symmetry, see Ref.\ \cite{GB}. The domain structure observed there
is a consequence of the high degree of symmetry of their potential.

The energy difference between embedded ANO and CC vortices can be calculated in a similar manner to that of ANO and twisted vortices. Close to the bifurcation, $\alpha\approx\alpha_{\rm b}$,
\begin{equation}
 \label{eq:EdiffOm0}
 E_{\rm ANO} - E \approx 2\pi (\alpha - \alpha_{\rm b}) \int r \d r f_2^2\,.
\end{equation}
According to Eq.\ (\ref{eq:EdiffOm0}), the energy of CC vortices is lower than that of embedded ANO vortices.

\begin{figure}
\centering
\subfigure[$\beta_1=\beta_2=\alpha=2$ and $\beta'=2.1$]{
\includegraphics[scale=.32,angle=-90]{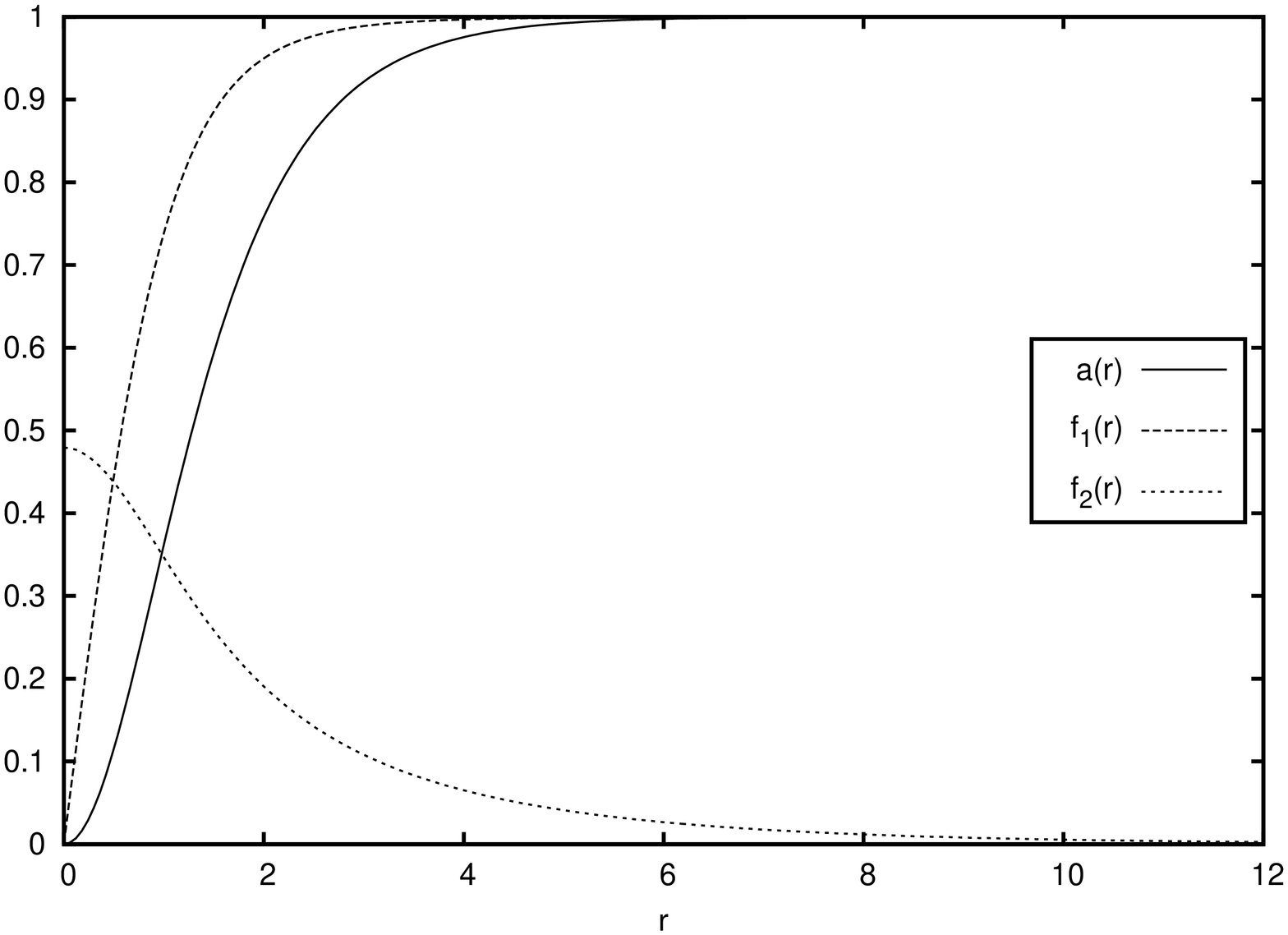}
\label{fig:vortom0}
}
\subfigure[$\beta_1=\beta'=\alpha=2$ and $\beta_2=3$]{
\noindent\hfil\includegraphics[scale=.32,angle=-90]{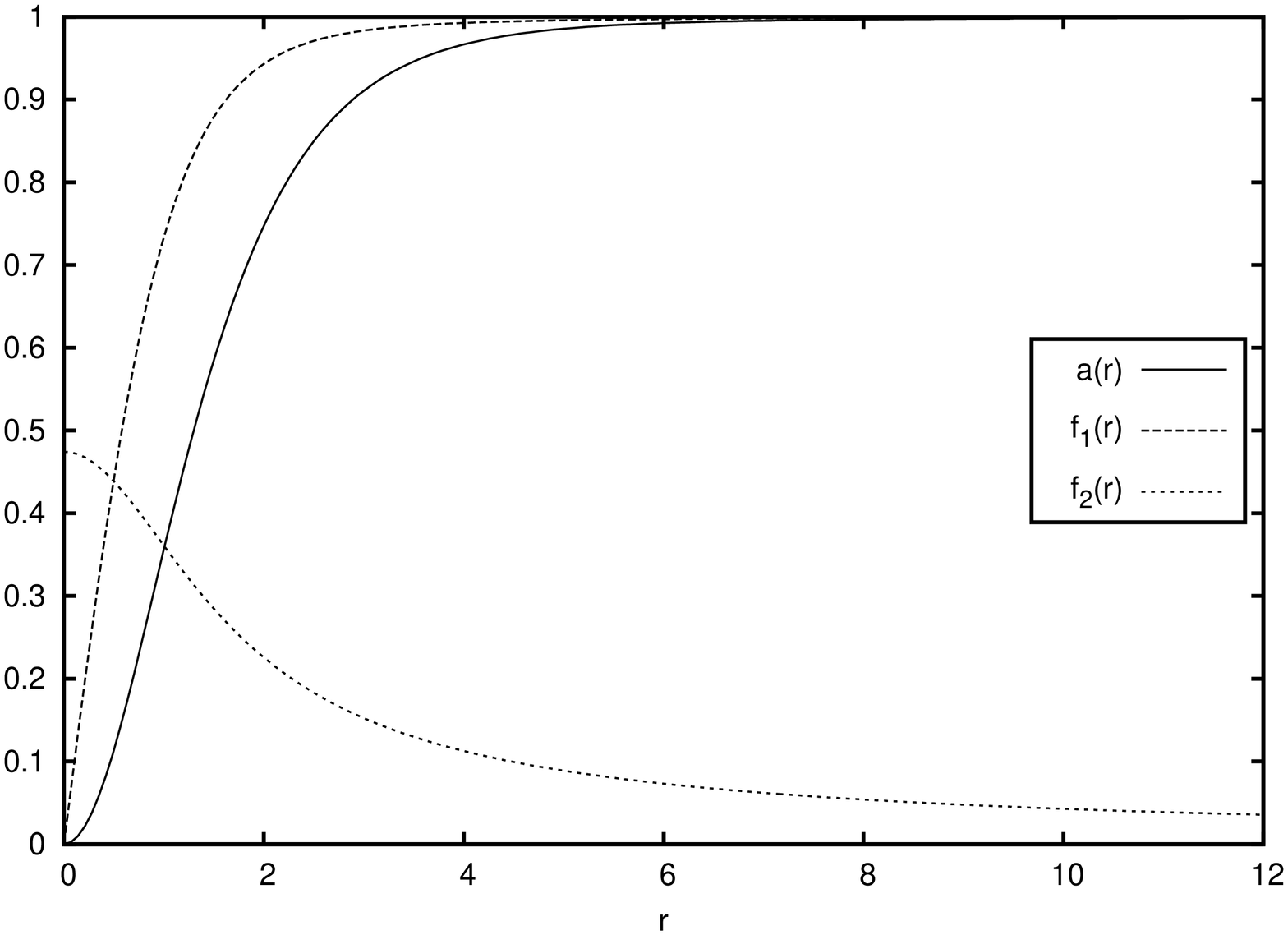}
\label{fig:vortom0P}
}
\caption{Zero twist solutions}
\end{figure}

\paragraph{Condensate core vortices, 2 charged fields} The zero twist limit of twisted vortices, condensate core vortices were calculated for a number of parameter values. One such solution, with exponential radial localisation (i.e., $\beta' > \alpha$) is shown in Fig.\ \ref{fig:vortom0}. In Fig.\ \ref{fig:vortom0P}, on the other hand, a CC vortex with power-law localisation is shown.
The energies of condensate core vortices are collected in Table \ref{tab:erg}. As already mentioned, their energies are below that of the embedded ANO vortex with the same value of $\beta_1$.

For ANO vortices for $\beta>1$, $E_n/n$ assumes its minimum for $n=1$, rendering higher flux vortices unstable. Interestingly, for CC vortices this is not the case. The minimum of $E_n/n$ is
assumed at a finite value of $n$. A plot of $E_n$ vs.\ $n$ is shown in Fig.\ \ref{fig:En}.

\begin{table}
 \begin{center}
 \begin{tabular}{|c||c|c|c||c|}
    \hline
    $n$            & (a)      & (b)        & (c)        & ANO\\
    \hline \hline
    1              & 1.152   & 1.008       & 0.78       & 1.157 \\
    2              & 1.121   & 0.913       & 0.75       & 1.210 \\
    3              & 1.107   & 0.882       & 0.72       & 1.239 \\
    \hline
 \end{tabular}
  \end{center}
  \caption{Energy per unit flux, $E_n/(2\pi n)$ of CC vortices for (a) $\beta_{1,2}=\alpha=2$, $\beta'=2.1$, (b) $\beta_1=2$, $\beta_2=8$, $\beta'=4.2$, $\alpha=4$
  and (c) $\beta_1=2$, $\beta_2=3872$, $\beta'=87.4$, $\alpha=83$
  compared to ANO $\beta=2$.}
  \label{tab:erg}
\end{table}

\paragraph{Condensate core vortices: A charged and a neutral field}
To obtain the range of parameters where solutions exist, we need to solve the bifurcation equation, Eq.\ (\ref{eq:bifur-f2}) again. Result for some parameter values are displayed
in Table \ref{tab:bif_e20}. Here, as the twist $\omega$ is obtained with a trivial transformation, we have collected $\alpha_{\rm b}$.
We have also calculated full nonlinear solutions numerically. Their energy values are shown in Table \ref{tab:erg2}. We would like to draw the attention to the fact, that $E_n/n$ is usually
a non-monotonous function of $n$, leading to stable higher flux vortices for $\beta_1 >1$. In these cases, embedded ANO vortices are unstable both against the formation of the condensate and against decay into unit flux vortices.

\begin{table}
 \centering
 \begin{tabular}{|c|c||c|}
  \hline
  $\beta_1$ & $\beta'$ & $\alpha_{\rm b}$ \\
  \hline\hline
  1.25      & 1.25     & 1.1235 \\
  2         & 2        & 1.7610 \\
  2.5       & 2.5      & 2.1791 \\
  1.25      & 1.255    & 1.1272 \\
  2         & 2.1      & 1.8309 \\
  2         & 2.3      & 1.9669 \\
  2         & 3.98372  & 2.9586 \\
  2.5       & 2.6      & 2.2477 \\
  \hline
 \end{tabular}
 \caption{The value of coupling $\alpha$ at the bifurcation for $e_2=0$.}\
 \label{tab:bif_e20}
\end{table}

\begin{table}
 \centering
 \begin{tabular}{|c||c|c||c|}
  \hline
  $n$  & (a)    & (b)    & ANO   \\
  \hline\hline
  1    & 1.152  & 1.113  & 1.157 \\
  2    & 1.104  & 1.054  & 1.210 \\
  3    & 1.102  & 1.011  & 1.239 \\
  \hline
 \end{tabular}
 \caption{Energy per unit flux, $E_n/(2\pi n)$, of vortices with $e_2=0$ for (a) $\beta_1=2$, $\beta_2=3$, $\beta'=2.3$ and $\alpha=2.05$ and
  (b) $\beta_1=2$, $\beta_2=9$, $\beta'=3.98372$, $\alpha=3.5507$.}
 \label{tab:erg2}
\end{table}

\section{Linear perturbations and stability}\label{sec:linpert}
To assess the stability of the solutions obtained, a linear stability analysis of the solutions has been performed. The formalism of Ref.\ \cite{Goodband} has been
used, extended to the case of two components. The $SU(2)$ symmetric case, has been considered in Ref.\ \cite{twistedinstab1, FL, twistedinstab2}.
Here, the Lagrangian of the theory, Eq.\ (\ref{eq:Lag}) is expanded to second order in small fluctuations of the fields, $\delta\phi_a$ and $\delta A_\mu$, and then, the resulting equations are solved with the help of a suitable form of partial wave expansion and Fourier transformation in $t,z$.
An important part of the procedure is the choice of gauge. The gauge condition is also perturbed,
in a way that removes first order derivatives from the first order equations \cite{Goodband}. The only drawback of this procedure is that the spectrum of the gauge fixing operator is also needed to distinguish physical modes from ghost ones, however, in our case, all ghost mode eigenvalues turn out to be positive, i.e., all unstable modes are physical.

The resulting equations, for a mode in partial wave channel $\ell$, and $z$ direction wave number $k$ can be written in the form
\begin{equation}
 \label{eq:genpert}
 \mathcal{M}_\ell(k) \Psi_\ell = \Omega^2 \Psi_\ell\,,
\end{equation}
where $\Omega$ is the frequency eigenvalue, $\mathcal{M}_\ell(k)$ a matrix differential operator, and $\Omega^2 <0$ corresponds to an instability.
Here $\Psi_\ell=(s_{1\ell}, s_{1,-\ell}^*, s_{2\ell}, s_{2,-\ell}^*, a_\ell, a_{-\ell}^*, a_{3\ell}, a_{0,\ell})^T$ are the radial functions of the perturbations.
For the details of this analysis, see Appendix \ref{app:pertdetails}. The perturbations of $a_{0,\ell}$ decouple in all cases.

The linearised problem, and its application to assess the stability of the solutions will be presented in Sec.\ \ref{ssec:1VEV11} for the case
of $e_1=e_2=1$. For the embedded ANO string, the following sectors of the perturbations decouple: $\delta\phi_1$, $\delta A_i$; $\delta A_0$; $\delta A_3$;
and that of $\delta \phi_2$. The instability in the $\delta\phi_2$ sector signals the bifurcation, the perturbation operator in that sector
agrees with that in the bifurcation equation, Eq.\ (\ref{eq:bifur-f2}).

The application of the expansion of the vortex solution around the bifurcation to the stability problem has been addressed in Ref.\ \cite{FL}
in the $SU(2)$ symmetric case.
The same argument can be repeated here, $\mathcal{M}_\ell = \mathcal{M}_\ell^{(0)} + \epsilon^2 \mathcal{M}_\ell^{(2)}$. This shows, that
the perturbation problem of the twisted vortices is a one-parameter deformation of that of the embedded ANO vortices, and therefore, twisted vortices
close to the bifurcation are unstable. Vortices farther from the bifurcation need to be treated numerically.

Let us also remark, that the perturbation treatment of the instability problem is a bit involved: for $\beta_1 > 1.5$, a contribution from the
continuum spectrum of the embedded ANO vortex perturbations (as intermediate states in 2nd order perturbation theory) needs to be taken into account \cite{FL}.

\subsection{Stability of vortices with two charged fields}\label{ssec:stabe21}
For twisted vortices, $0 < \omega \le \omega_{\rm b}$, the results are similar to those in the case of an $SU(2)$ symmetric potential (see Refs.\ \cite{twistedinstab1, FL,twistedinstab2}):
fistly, the mode corresponding to the lovest value of the squared frequency $\Omega^2$ is a one-parameter deformation of the instability mode of the embedded ANO vortex.
Second, for all values $0<\omega \le \omega_{\rm b}$ which were available to our numerical code, we have found one unstable mode in the $\ell=0$ sector, i.e.,
the instability of the embedded ANO vortex persisted for all examined twisted vortices, and, for lower values of the twist, $\omega$, the value of
$|\Omega^2|$ got also smaller. The value of $\Omega^2$ is negative for a range of the wave number. Close to the minimum $k=k_{\rm min}$ (most negative $\Omega^2$), an approximate dispersion relation
\begin{equation}
  \label{eq:disprel}
  \Omega^2 = \Omega^2_{\rm min} +\Omega^2_2(k-k_{\rm min})^2\,,
\end{equation}
holds. For the embedded ANO vortex, Eq.\ (\ref{eq:disprel}) is exact, and $k_{\rm min} = \omega_{\rm b}$. Some data is displayed in Table \ref{tab:instab1}--\ref{tab:instab3}.
As $\omega$ becomes smaller, the errors grow; it is likely that this is because of $\delta A_3$ decoupling at $\omega=0$. For $\omega\to0$, a very small $\delta A_3$ has to be
calculated, which is weakly coupled to the other components. On the other hand, for $\omega=0$, the eigenvalues for $\delta A_3$ are those of the ghost mode (see Table \ref{tab:ghost}).

The ghost mode eigenvalues, collected in Table \ref{tab:ghost} and \ref{tab:ghost2}. They change slowly with parameters of the potential, and $\omega$. Their order of magnitude is
unity. Most importantly, the lowest energy modes relevant for stability are not cancelled by them.

Most importantly for our subject matter, in all examined cases, the eigenvalues in the $\omega=0$ case are $0$ for $k=0$ within numerical precision. For zero twist ($\omega=0$),
the dispersion relation (\ref{eq:disprel}) is exact with $k_{\rm min}=0$ and $\Omega_2^2=1$. This implies, that for any $k\ne 0$, the eigenvalue is positive.
As any local perturbation necessarily contains modes with $k\ne 0$, it is a positive energy perturbation. This is a strong evidence for the stability of the zero twist vortices.

We have also examined the stability of higher flux vortices. We have found, that for many parameter values, $n=2,3$ vortices are stabilised by the addition of the condensate in their core.
This is in accord with the non-monotonicity of the energy per unit flux as a function of the flux. For vortices with number of flux quanta below the one with the strongest binding, it is energetically favourable to avoid decay. This happens when the parameters are far enough from the bifurcational value. See Table \ref{tab:stabe21}: CC vortices exists for $\alpha_{\rm b} < \alpha < \beta'$,
and they are stable for $\alpha > \alpha_{\rm s}$. As we shall see in the large mass ratio limit, in Sec.\ \ref{sec:limits}, this phenomenon is even more pronounced.

\begin{table}
 \centering
 \begin{tabular}{|c|c|c||c|c|}
  \hline
  $\beta_1$  & $\beta_2$  & $\beta'$  & $\alpha_{\rm b}$ & $\alpha_{\rm s}$ \\
  \hline\hline
  1.25       & 1.25       & 1.255     & 1.1742           & 1.2350           \\
  2          & 2          & 2.1       & 1.6811           & 1.9335           \\
  2.5        & 2.5        & 2.6       & 1.9756           & 2.3984           \\
  \hline
 \end{tabular}
 \caption{Stabilisation of two-flux ($n=2$) vortices, $e_2=1$.}
 \label{tab:stabe21}
\end{table}

\subsection{Stability of vortices with one charged and one neutral fields}\label{ssec:stabe20}
We have also examined the stability of CC vortices in the $e_2=0$ case. We have found qualitatively similar results, as in the charged case: the eigenvalue of the mode that is a deformation of the eigenmode of ANO vortices corresponding to the bifurcation looses its energy lowering property for CC vortices. The corresponding eigenvalue becomes zero within numerical precision, for $z$-independent perturbations,
and $k^2$ for perturbations with $z$ direction wave number $k$, implicating that there are no energy-lowering local perturbations in this sector.

For higher winding vortices, we have also observed the stabilisation in the case of a neutral second field. For some numerical data, see Tab.\ \ref{tab:stabe20}.

\begin{table}
 \centering
 \begin{tabular}{|c|c|c||c|c|}
  \hline
  $\beta_1$  & $\beta_2$  & $\beta'$  & $\alpha_{\rm b}$ & $\alpha_{\rm s}$ \\
  \hline\hline
  2          & 3          & 2.3       & 1.4325           & 1.8287           \\
  2          & 9          & 3.98372   & 1.9448           & 2.7938           \\
  \hline
 \end{tabular}
 \caption{Stabilisation of two-flux ($n=2$) vortices, $e_2=0$.}
 \label{tab:stabe20}
\end{table}

\section{Magnetic bags and large mass ratio \texorpdfstring{$M$}{M}}\label{sec:limits}

\begin{figure}[h!]
 \noindent\hfil\includegraphics[angle=-90,scale=.5]{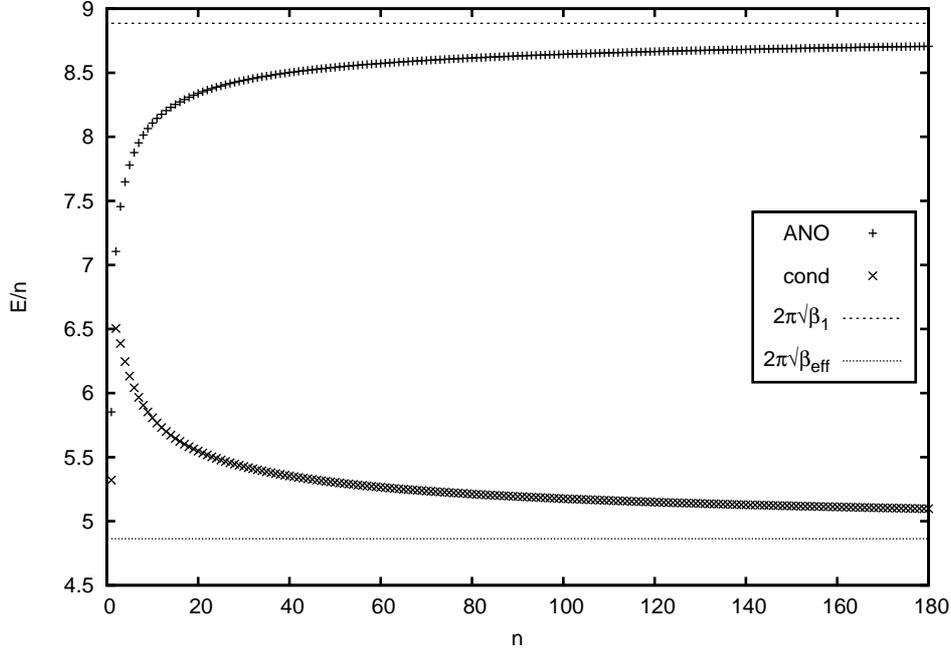}
 \caption{Energy of vortices per unit flux, $\beta_1=2$, $\tilde{\beta}_2=3$, ${\tilde{\beta}}'=2.3$, $\tilde{\alpha}=2.05$ and $e_2=0$,
 compared to Abrikosov (ANO) vortex energies.}
 \label{fig:Ergs}
\end{figure}

\begin{figure}[h!]
 \noindent\hfil\includegraphics[angle=-90,scale=.5]{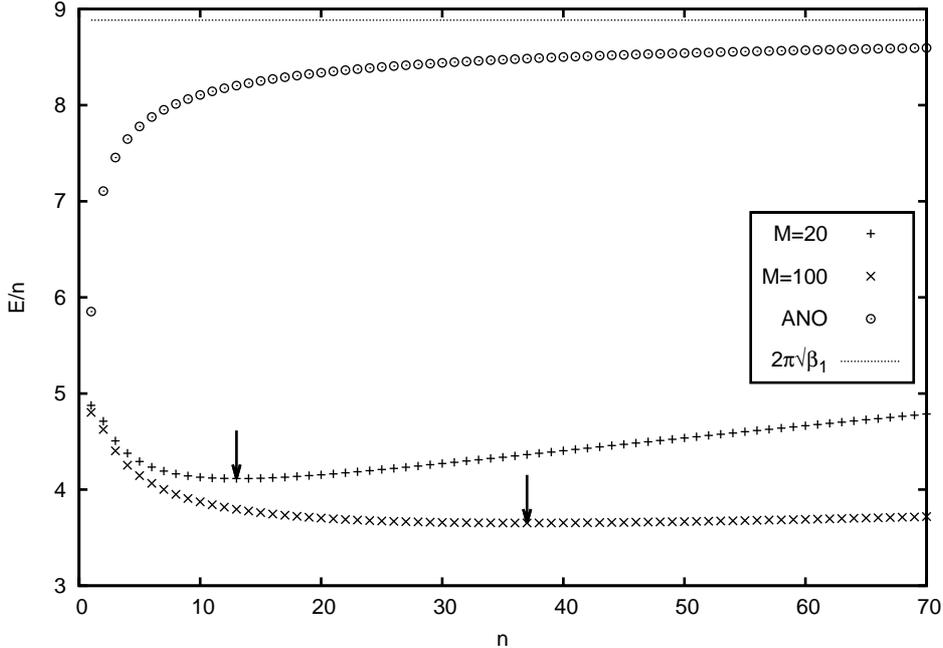}
 \caption{Energy of vortices per unit flux, $\beta_1=2$, $\beta_2=M^2\tilde{\beta}_2$, $\beta'=M {\tilde{\beta}}'$, $\alpha=M\tilde{\alpha}$, $\tilde{\beta}_2=9.68$,  ${\tilde{\beta}}'=4.37$, $\tilde{\alpha}=4.15$ and $e_2=1$,  compared to Abrikosov (ANO) vortex energies. The arrows show the minima, at $n=13$ and $n=78$, respectively.}
 \label{fig:En}
\end{figure}

\paragraph{Large flux} A remarkable limit of ANO vortices has been considered in Ref.\ \cite{Bolognesi1, Bolognesi2}. An approximate vortex  configuration has been constructed as
\begin{equation}
 \label{eq:Bolognesi}
 f(r) = \left\{ \begin{aligned}
                 0 &\text{, if } r < R\,,\\
                 1 &\text{, if } r > R\,,\\
                \end{aligned}
\right. \quad a(r) = \left\{ \begin{aligned}
                              r^2/R^2 &\text{, if } r < R\,,\\
                              1 &\text{, if } r > R\,,\\
                             \end{aligned}
\right.
\end{equation}
with optimal radius $R=R_A=\sqrt{2 n}\beta^{-1/4}$ and energy $E_n = E_{An} \sim 2\pi n \sqrt{\beta}$. It is straighforward to generalise this approximation to the case of a neutral second field,
$e_2=0$ with using Eq.\ (\ref{eq:Bolognesi}) for $f_1$ and $a$, and setting
\begin{equation}
 f_2(r) = \left\{ \begin{aligned}
                   \sqrt{\frac{\alpha}{\beta_2}} &\text{, if } r < R\,,\\
                   0 &\text{, if } r > R\,,
                  \end{aligned}\right.
\end{equation}
yielding $R = R_{C0} = \sqrt{2 n}\left( \beta_1 -\alpha^2/\beta_2\right)^{-1/4}$ and $E=E_{C0}= 2\pi n \left( \beta_1 -\alpha^2/\beta_2\right)^{1/2}$. It is remarkable, that in this limit, an effective
Ginzburg-Landau parameter, $\beta_{\rm eff}=\beta_1 - \alpha^2/\beta_2$ can be introduced.
In the $e_2=0$ limit, the large flux limit of the effective ANO vortex reproduces well the large flux limit of CC vortices as well.
However, the large flux behaviour is more delicate in the case of two charged fields: in that case, we have observed numerically, that for $n\to\infty$, $E_n/n$ approaches the same limit  for CC and ANO vortices. For numerical data, see Figs.\ \ref{fig:En} and \ref{fig:Ergs}.

\paragraph{Large mass ratio, $M$}
The large mass ratio limit is another interesting limit, and one that is also physically relevant. In LMH, $\phi_1$ corresponds to Cooper pairs formed of electrons, and $\phi_2$ to ones of protons.
The GL free energy density is
\begin{equation}
 \label{eq:GL}
 \mathcal{F} = \frac{{\mathbf B}^2}{2} + \sum_{a=1}^2 \left[ \frac{\hbar^2}{2 m_a}|\mathbf{D}\phi_a| + \frac{\lambda_a}{2}|\phi_a|^4 - \mu_a |\phi_a|^2 \right] + \lambda' |\phi_1|^2 |\phi_2|^2\,,
\end{equation}
where ${\mathbf D}\phi_a = (\nabla - e e_a {\mathbf A})\phi_a$, $\lambda_a$, $\lambda'$, and $\mu_a$ are material constants, $e_a$ is the charge of the field $\phi_a$ in some arbitrary units $e$ (e.g., for superconductors, twice the elementary charge is suitable), and we have assumed that there is no Josephson coupling, $\gamma (\phi_1^*\phi_2 + \phi_1 \phi_2^*)$, which would fix the relative phase of the fields at the minimum of the potential, and disallow a 1VEV state. Such is the case if there is a symmetry enforcing the separate conservation of the two fields (e.g., conservation of particle numbers).

With the help of a rescaling of the field $\phi_1$ by $\sqrt{\mu_1/\lambda_1}$, $\phi_2$ by $\sqrt{m_2/m_1} \sqrt{\mu_1/\lambda_1}$, the vector potential ${\bf A}$ by $\hbar\eta_1 \sqrt{\mu_1/2/m_1}$ and distances by $\sqrt{2m_1/(\mu_0e_1^2e^2\eta_1)}$, the penetration depth, $\lambda_L = \sqrt{m_1/(\mu_0e_1^2e^2\eta_1)}$ is scaled to $1/\sqrt{2}$, and we obtain the GL free energy with the potential in the form used in Eq.\ \ref{eq:pot}. The parameters are then related to the microscopic parameters as
\begin{equation}
 \label{eq:supercond}
 \begin{aligned}
    \beta_1 &= 4\lambda_1 m_1^2/(\hbar^2e^2\mu_0)\,,\\
    \beta'  &= 4 \lambda' m_1 m_2/(\hbar^2 e^2 \mu_0)\,, \\
 \end{aligned}
 \quad\quad
  \begin{aligned}
   \beta_2 &= 4\lambda_2 m_2^2/(\hbar^2 e^2\mu_0) \,,\\
   \alpha  &= 4 \nu_2 m_1 m_2 /(\hbar^2 e^2 \mu_0 \eta_1^2)\,.\\
  \end{aligned}
\end{equation}
As for LMH, the mass ratio, $M=m_2/m_1 \approx 1836$ (the mass ratio of protons and electrons), $\beta_2 \gg \alpha, \beta' \gg \beta_1$. Suitable parameters are introduced as
\begin{equation}
 \label{eq:scaledparams}
 \beta_2 = M^2 \tilde{\beta}_2\,,\quad \beta' = M {\tilde{\beta}}'\,,\quad \alpha = M\tilde{\alpha}\,.
\end{equation}
The tilde parameters are expected to be of the same order of magnitude.
We shall consider here the limiting behaviour of CC vortices for $M\gg 1$.

In Fig.\ \ref{fig:En}, the energy per unit flux, $E_n/n$ is plotted for a large range of fluxes, and some values of the mass ratio. Remarkably, $E_n/n$
is not a monotonous function of $n$, in contradistinction with both the case of type I ($\beta < 1$) and type II ($\beta > 1$) superconductors. As a result, in the two component theory with large mass ratio, ``giant'' vortices exist. Even for moderate values of $M$ (e.g., 20 or 100), the minimum of $E_n/n$ is shifted to 13, resp.\ 78.

Qualitative properties of the function $E_n/n$ can be reproduced with the following approximate vortex configuration.
Let ud consider a bag-type vortex, with $f_1=0$, $f_2=\sqrt{\alpha/\beta_2}$ (the lowest energy false vacuum with $f_1=0$) in its core,
from $r=0$ to $(1-\delta)R$. It is assumed that the vortex has a thin wall, with $f_1$ and $f_2$ hhaving a linear transition to their respective
VEVs between $r=(1-\delta)R$ and $R$. The gauge field is $a=(r/R)^2$ for $r<R$ and $a=1$ otherwise.

The energy of such a configuration is approximately minimised is $R=\sqrt{2 n}\beta_{\rm eff}^{-1/4}$. In $\delta$ we expad the energy in a series containing terms starting with $1/\delta$ and ending with $\delta^3$. We have found, with a numerical minimisation, that a good approximate minimum is obtained by minimising the $\delta^{-1}$ and $\delta^3$ terms, yielding $\delta=(5/2)^{-1/4}((\beta_2 +\alpha)/(\beta_2-3\alpha))^{1/4}n^{-1/2}$. With these, it is obtained that
\begin{equation}\label{eq:ErgApprnM}\begin{aligned}
 E \approx 2\pi n \beta_{\rm eff} &+ \frac{8\pi}{3}\left( \frac{2}{5}\right)^{1/4}\left[ 1
-\frac{1}{4\sqrt{10}}\left( 7\beta_{\rm eff} - \frac{\tilde{\alpha}(\tilde{\alpha}+{\tilde{\beta}}')}{\beta_{\rm eff}\tilde{\beta}_2} \right)
 \right] n^{1/2}\\
 &+\frac{\pi \tilde{\alpha}}{M\tilde{\beta}_2} \left[ 1 - \frac{2^{7/4}5^{1/4}}{3}n^{3/2} + \frac{3 \cdot 5^{1/2}}{2^{1/2}}n \right]\,.
\end{aligned}
\end{equation}
The qualitative formula (\ref{eq:ErgApprnM}) gives an order of magnitude correct value. It also shows, that $E_n/n$ is nonmonotonous, with a minimum
at a value of $n$ growing with $M$. This minimum is significantly below the energy/flux of embedded ANO vortices (in the bag approximation of Refs.\ \cite{Bolognesi1, Bolognesi2}, $2\pi \beta_1$).
The existence of the minimum is the result of the competition of two phenomena, the expansion of the vortices due to the magnetic energy, and the
large $M$ behaviour, fixing $f_2$ to its minimal energy value in the core, at the cost of the interaction energy between the second scalar and the gauge fields.  If $n$ becomes much larger than at the minimum of $E_n/n$, CC vortices approach embedded ANO ones.

\paragraph{Boundary of upper and lower component 1VEV: Wall-type vortices} Close to $\alpha=\sqrt{\beta_1 \beta_2}$, the potential energy in the core becomes small, the vortices become large, and their flux is localised closer to
the outer end of their cores. At the same time, the minimum of $E_n/n$ is shifted to larger values of $n$, and at $\alpha=\sqrt{\beta_1 \beta_2}$,
$E_n \propto n$ for large $n$. Here, ANO vortices in the lower component become also allowed.
In this case, it is possible to exchange the role of the 2 components, with the rescaling $\phi_a\to \eta_2\phi_a$, $x\to x/\eta_2$, $A\to \eta_2 A$, where $\eta_2^2=\alpha/\beta_2$.
In this way, we get the same expression for the energy of the vortices with the potential (\ref{eq:pot}) and an overall multiplier $\alpha/\beta_2$.
With the same configuration as above, the estimated energy of these vortices is $E=2\pi( 4\alpha/\beta_2 + \alpha/\sqrt{3\beta_2})$, which
is $M^0$ asymptotically. However, using the large-$\beta$ asymptotics of Abrikosov vortex energy \cite{Pismen}, we get $E\sim 2\pi\frac{\alpha}{\beta_2}\log\sqrt{\beta_2}$,
i.e., $\sim (\log M)/M$, telling us that at the transition, it is energetically favourable for the vortices to break up into $n=1$ lower component
Abrikosov vortices. Linearising the equations in the other component shows, that these vortices are then stable against the formation of a condensate
in their core. This can be seen as follows: the large-$\beta$ asymptotic form of the vortex profile is a small core
with size proportional to $1/\sqrt{\beta_2}\propto 1/M$. The linearised equation is of the form of an eigenvalue equation,
and we have verified numerically, that it has no bound modes, and therefore if $\alpha > \sqrt{\beta_1 \beta_2}$, vortices in the lower component do not have condensate in their cores.

\section{The case of a two-component vacuum expectation value}\label{sec:2VEV}
\paragraph{Global 2VEV vortices} Let us briefly consider global 2VEV vortices. These, in the context of atomic BECs, are discussed in Refs.\ \cite{Kasamatsu, Mason, IvashinPoluektov, KasamatsuEtoNitta}. Let us note, that as for $r\to\infty$, $f_1\to \eta_1$ and $f_2\to\eta_2$, the asymptotic behaviour of the energy density [see Eq.\ (\ref{eq:globerg})] is $\mathcal{E} \sim (n^2 \eta_1^2 + m^2 \eta_2^2)/r^2$, therefore, the energy of the vortex is
\begin{equation}
 \label{eq:2VEVglobErg}
 E = \int \d^2 x \mathcal{E} = 2\pi\int_0^{R_{\rm core}} \d r r \mathcal{E} + 2\pi (n^2 \eta_1^2 + m^2 \eta_2^2) \log\left( \frac{R}{R_{\rm core}}\right)\,.
\end{equation}

\begin{figure}
\centering
\includegraphics[scale=.5,angle=-90]{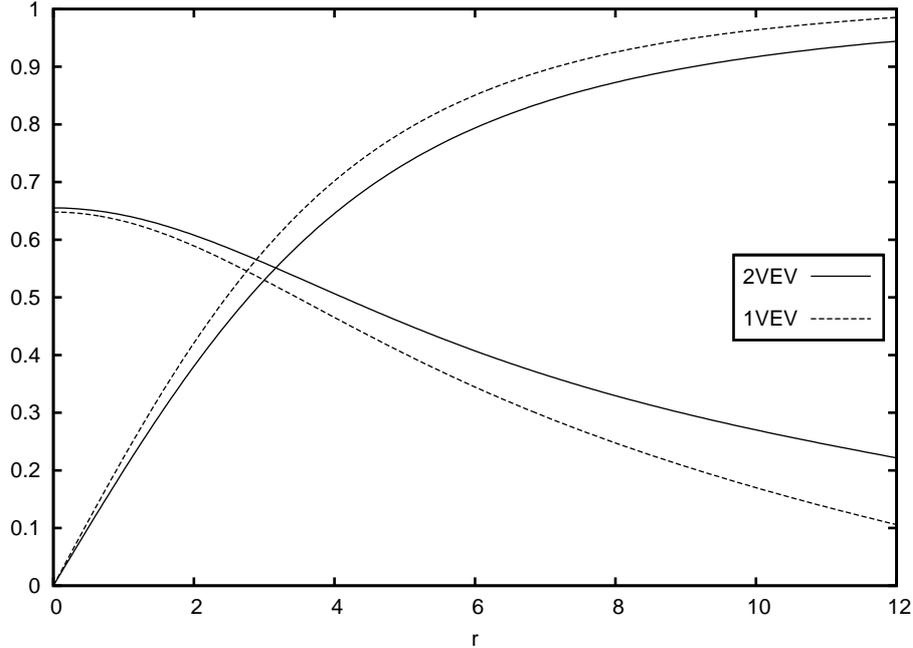}
\caption{Comparison of 1VEV and 2VEV global vortices close to the boundary: $\beta_1=\beta'=2$, $\beta_2=4.5$, $\alpha=1.99$ 1VEV and $\alpha=2.011$ 2VEV.}
\label{fig:2vevcmpglob}
\end{figure}

An interesting case is the behaviour close to the boundary between 1VEV and 2VEV classes, at $\alpha=\beta'$. Unless $\beta_1 \beta_2 = (\beta')^2$,
a limiting vortex exists in the 1VEV case, with power-law localisation. It is also a smooth limit of 2VEV vortices: at the transition, $\eta_2$ becomes 0.
For the comparison of an 1VEV and a 2VEV global vortex, both close to the transition, see Fig.\ \ref{fig:2vevcmpglob}. Numerical data is collected in Table \ref{tab:trans}.

\paragraph{Two charged fields} Let us note first, that with two non-zero VEVs, the energy per unit length of a twisted vortex diverges quadratically in $R$,
as there is only one longitudinal gauge field component, $A_3$, which would either not cancel the longitudinal derivative of $\phi_2$, or lead to a non-vanishing $D_3 \phi_1$.
Also, as $A_\vartheta$ cannot cancel the angular derivatives of both fields, unless $n=m$, the energy of 2VEV vortices is only finite in this case.

Vortices with a mildly, i.e., logarithmically divergent energy exist, however, for any pair of windings, $n$, $m$. From minimising the logarithmic energy contribution, $a(r\to\infty) = \eta_1^2/(\eta_1^2 + \eta_2^2)$, agreeing with the number of flux quanta in the vortices, is obtained; in general, this is non-integer. In Refs.\ \cite{BabaevF, BS, BS1, BS2}, these vortices have been termed fractional flux vortices.

\begin{figure}
\centering
\includegraphics[scale=.5,angle=-90]{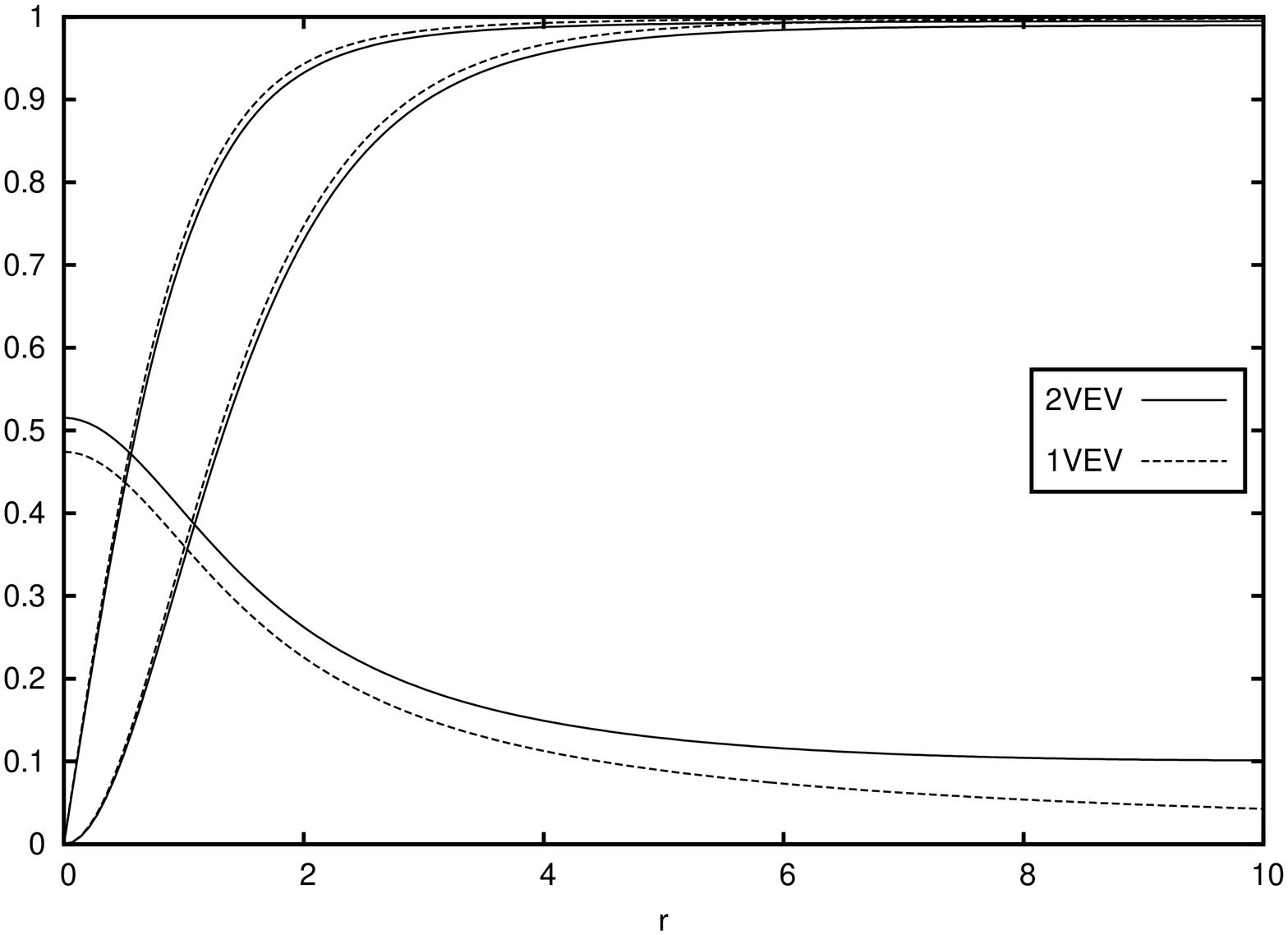}
\caption{Comparison of 1VEV and 2VEV vortices close to the boundary: $\beta_1=\beta'=\alpha=2$, $\beta_2=3$ 1VEV and $\alpha=2.011$ 2VEV, $e_2=1$.}
\label{fig:2vevcmpe21}
\end{figure}

Let us now consider the case of $n=1$, $m=0$. Inserting the limiting value of $a$ and the VEVs into the energy density, Eq.\ \ref{eq:Edens}, the asymptotic form of the energy is obtained, yielding
\begin{equation}
 \label{eq:2VEVdivergence}
 E = 2\pi \int_0^{R_{\rm core}} \d r r \mathcal{E} + 2 \pi E_L \log\left( \frac{R}{R_{\rm core}}\right) = E_{\rm core} + 2 \pi E_L \log\left( \frac{R}{R_{\rm core}}\right)\,,
\end{equation}
where the coefficient of the logarithm is given as
\begin{equation}\label{eq:2VEVchgEL}
 E_L = \frac{\eta_1^2 \eta_2^2}{\eta_1^2 + \eta_2^2}\,.
 \end{equation}
In the 1VEV case, close to the transition, the radial fall-off of the second field component is $\sim F_2 r^{-1/2}\exp(-\sqrt{\beta' -\alpha}r)$, which gets slower if the system is closer to the 2VEV case. For a finite size sample, at some point, 1VEV solutions and 2VEV fractional flux vortices become indistinguishable in those cases when the zero twist limit exists for $\beta'=\alpha$. For a comparison of 1VEV and 2VEV vortices close to the transition, see Figs.\ \ref{fig:2vevcmpe21}, and Table \ref{tab:trans}.

Let us also mention, that in the large mass ratio ($M$) limit, $\eta_1$ is independent of $M$, and $\eta_2^2 = {\tilde{\eta}}_2^2 / M$. As a result, in the large mass ratio limit, $E_L = O(M^{-1})$, and the dependence of the energy on $R$ becomes weak. The limit of the flux is
\[
n a(r\to\infty)= n \frac{\eta_1^2}{\eta_1^2 + \eta_2^2} =
n \left[ 1 - \frac{1}{M}\frac{\beta_1 (\tilde{\alpha}-{\tilde{\beta}}')}{\beta_1 \tilde{\beta}_2 -\tilde{\alpha}{\tilde{\beta}}'}\right] + O(1/M^2)\,,
\]
 i.e., the deviation of the flux from the integer value in the $M\gg 1$ limit decreased with $M$, and in the case of LMH,
distinguishing between fractional flux and ANO vortices is expected to require the measurement of the flux to a precision of
less than one part in a thousand. At the same time, the coefficient of the logarithmic term [see Eq.\ (\ref{eq:2VEVchgEL})] in the energy also becomes small,
\[
E_L = \frac{1}{M} \frac{\beta_2 (\tilde{\alpha}-{\tilde{\beta}}')}{\beta_1 \tilde{\beta}_2 - {\tilde{\beta}}'{}^2} + O(1/M^2)\,.
\]

\begin{figure}[h!]
 \noindent\hfil\includegraphics[angle=-90,scale=.5]{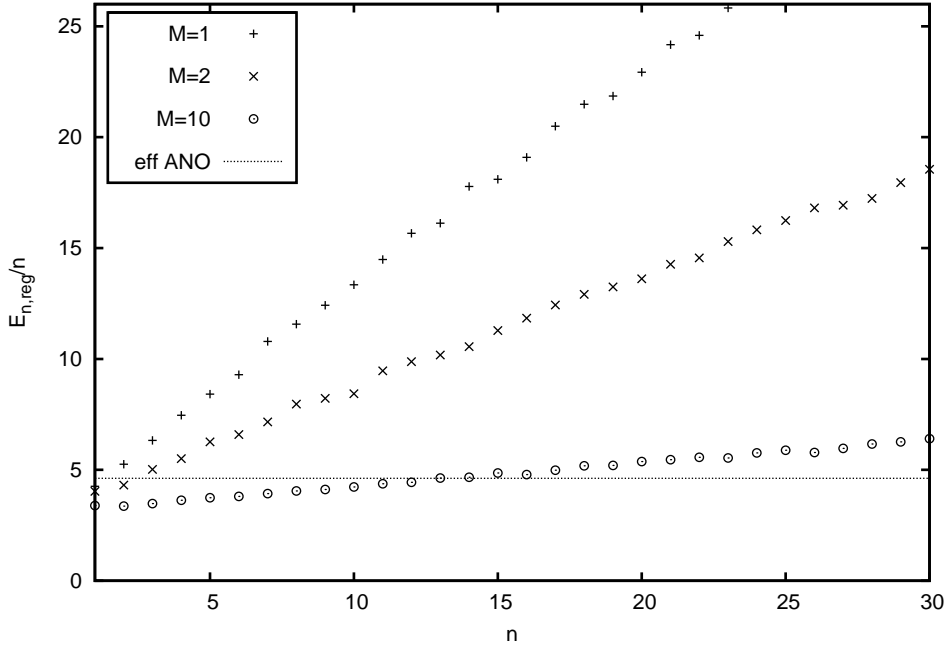}
 \caption{Energy of 2VEV vortices per unit flux, $\beta_1=2$, $\tilde{\beta}_2=2.4$, ${\tilde{\beta}}'=1.8$, $\tilde{\alpha}=2.2$ and $e_2=1$.
 For comparison, the energy per unit flux of the corresponding effective ANO vortices for large flux, $(\alpha_{\rm eff}/\lambda_{\rm eff}) 2\pi \sqrt{\lambda_{\rm eff}}$, is also indicated.}
 \label{fig:En2V1}
\end{figure}

\begin{figure}[h!]
 \noindent\hfil\includegraphics[angle=-90,scale=.5]{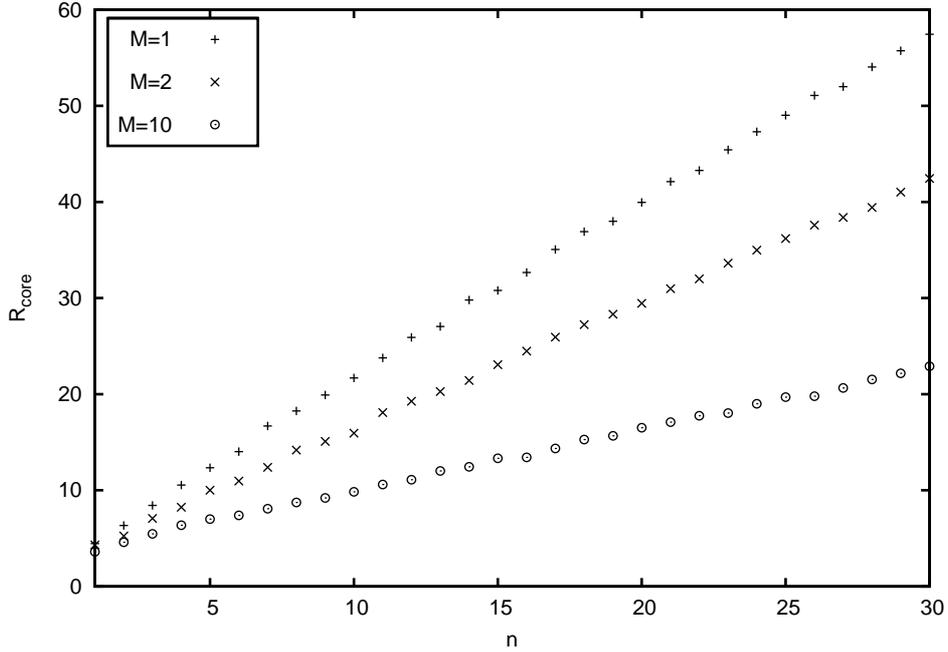}
 \caption{Radius of 2VEV vortices ($f_1(R_{\rm core})=0.95 \eta_1$), $\beta_1=2$, $\tilde{\beta}_2=2.4$, ${\tilde{\beta}}'=1.8$, $\tilde{\alpha}=2.2$ and $e_2=1$.}
 \label{fig:Rc2V1}
\end{figure}

Also, $f_1$ becomes similar close to the scalar field of an ANO vortex, with $\lambda_{\rm eff}$ and $\alpha_{\rm eff}$ effective couplings, as in the 1VEV case, see Sec.\ \ref{sec:limits}.
Some numerically calculated (core) energy values of 2VEV vortices with both fields charged are shown in Fig.\ \ref{fig:En2V1}, and the corresponding radii in Fig.\ \ref{fig:Rc2V1}. Note, that there seems to be an energy contribution proportional to $n^2$, and $E_n/n$ grows with $n$ despite that $\beta_{\rm eff} <1$.

\paragraph{One field charged, one neutral}
In the 2VEV case with one neutral condensate, $e_2=0$, the $m=0$ case yields finite energy vortices with integer flux, $a(r\to\infty) \to 1$, and the number of flux quanta agrees with $n$. For other values of $m$, $E \sim 2 \pi m^2 \eta_2^2 \log(R/R_{\rm core})$.

We have calculated some 2VEV vortices with $n=1$, $m=0$ for a neutral scalar field numerically. The data is collected in Table \ref{tab:2VEVe20}.
Note, that there is a series of data for $\beta_1=2$, $\tilde{\beta}_2=3$, ${\tilde{\beta}}'=2$, $\tilde{\alpha}=2.1$ and $M=1,2,3$. For $M\to\infty$,
the lowest energy state with $\phi_1=0$ is $|\phi_2|=\sqrt{\alpha/\beta2}=O(1/\sqrt{M})$. With this assumption, the leading terms in the equation of $f_2$
in Eq.\ (\ref{eq:FRVprofE}) is $\beta_2 f_2^2 -\alpha + \beta' f_1^2$, neglecting the remaining terms yields $f_2^2 = (\tilde{\alpha} - {\tilde{\beta}}'f_1^2)/\tilde{\beta}_2 / M$. Substituting this into the equation of $f_1$ yields an ANO vortex profile equation with $\lambda_{\rm eff}=\beta_1 - ({\tilde\beta}')^2/\tilde{\beta}_2$ and $\alpha_{\rm eff}=\beta_1 - {\tilde\beta}' \tilde{\alpha}/\tilde{\beta}_2$. Rescaling this into the usual ANO form yields an approximate energy $\alpha_{\rm eff}/\lambda_{\rm eff} E_{ANO}(\beta=\lambda_{\rm eff})$. For comparison, for the case in Table \ref{tab:2VEVe20}, this yields $E/(2\pi)\approx 0.8279$ (with $\lambda_{\rm eff}=0.6667$, $\alpha_{\rm eff}=0.6$, and $E_{ANO}(\beta=0.6667)/(2\pi)=0.9199$).

\begin{figure}[h!]
 \noindent\hfil\includegraphics[angle=-90,scale=.5]{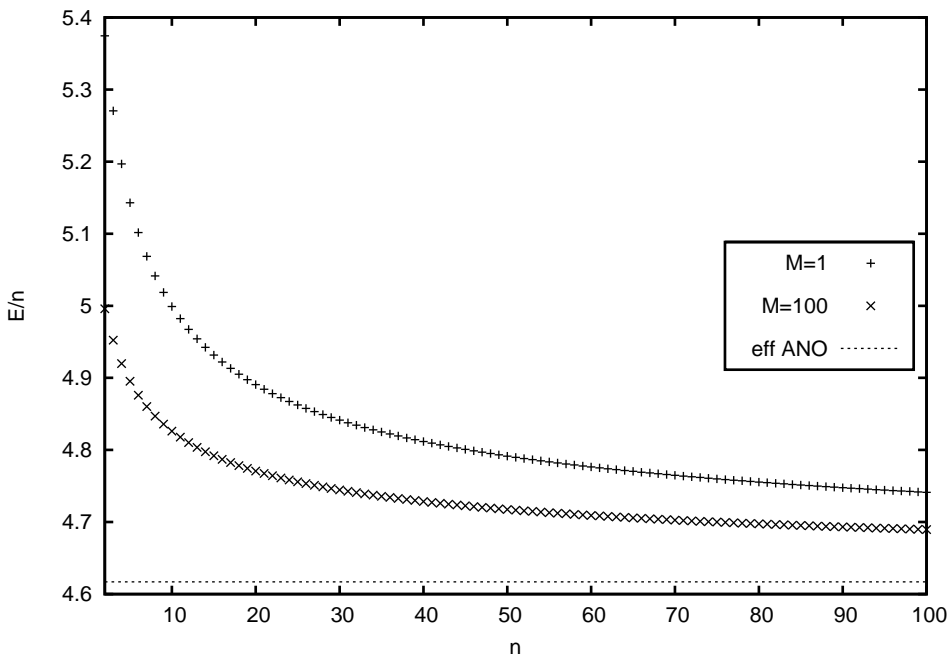}
 \caption{Energy of 2VEV vortices per unit flux, $\beta_1=2$, $\tilde{\beta}_2=3$, ${\tilde{\beta}}'=2$, $\tilde{\alpha}=2.1$ and $e_2=0$.
 For comparison, the energy per unit flux of the corresponding effective ANO vortices for large flux, $(\alpha_{\rm eff}/\lambda_{\rm eff}) 2\pi \sqrt{\lambda_{\rm eff}}$, is also indicated.}
 \label{fig:En2V0}
\end{figure}

Some data for $n=2$ is collected in Table \ref{tab:2VEVe202}. For $n=2$, the approximation from the effective ANO vortex gives
$E/(4\pi) \approx 0.8071$   (with $E_{ANO}(n=2,\beta=0.6667)/(4\pi)=0.8967$). See also Fig.\ \ref{fig:En2V0}.

Numerically, $f_2^2 \approx (\alpha - \beta' f_1^2)/\beta_2$ holds with a good accuracy even for $M=4$.

\begin{figure}
\centering
\includegraphics[scale=.5,angle=-90]{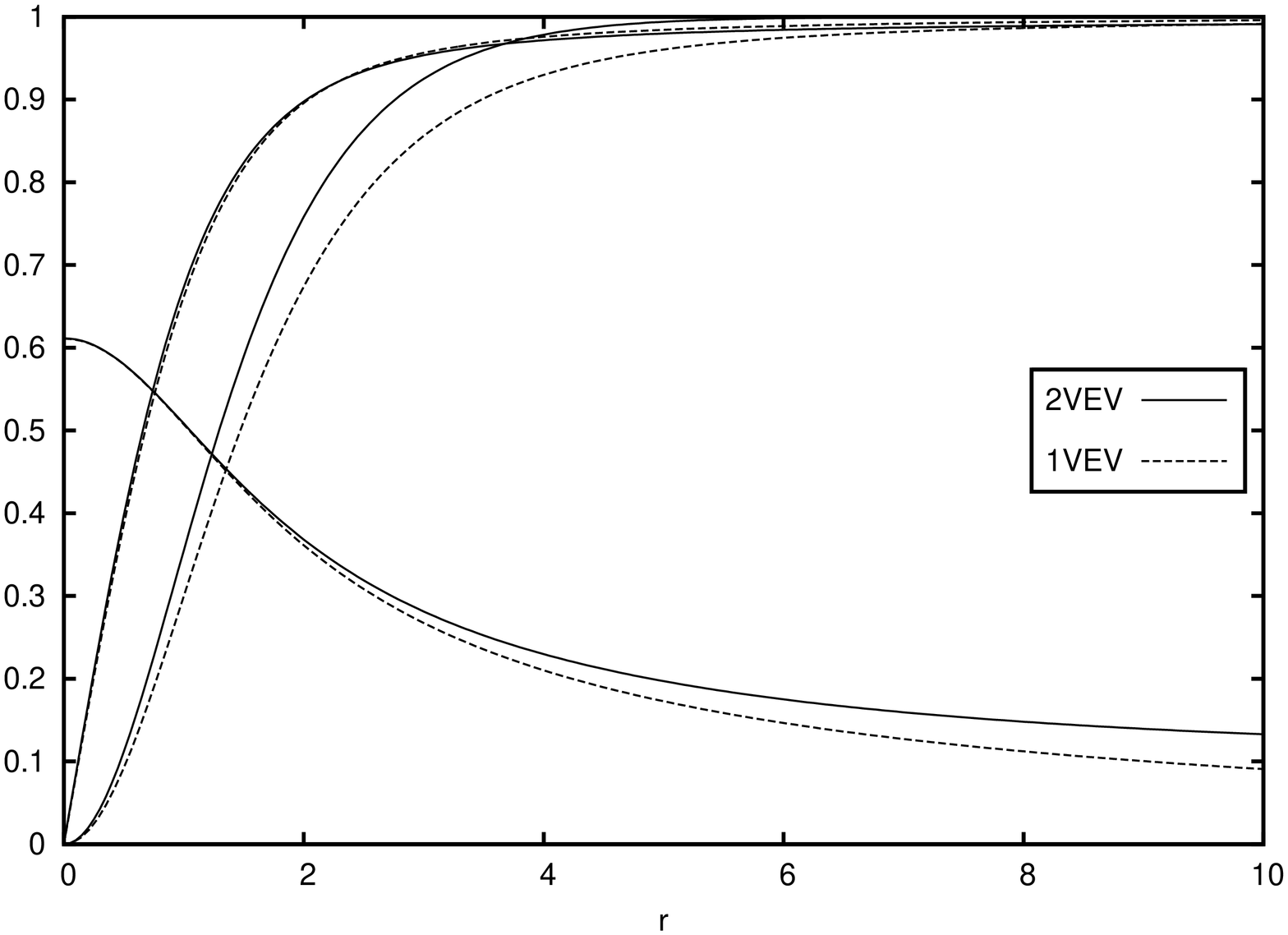}
\label{fig:2vevcmpe20}
\caption{Comparison of 1VEV and 2VEV vortices close to the boundary: $\beta_1=\beta'=\alpha=2$, $\beta_2=3$ 1VEV and $\alpha=2.011$ 2VEV, $e_2=0$.}
\end{figure}

\begin{table}
 \centering
 \begin{tabular}{|c|c| c | c || c | c | c || c| c | c|}
 \hline
      & $\beta_1$ & $\beta_2$ &  $\beta'$  & $\alpha$ & $E/(2\pi)$, 1VEV & $R_{\rm core}$  & $E_L$ & $E/(2\pi)$, 2VEV & $R_{\rm core}$ \\
 \hline\hline
 (a)  & 2         & 3         &  2         & 2        & 1.13598          & ---    & 0.010879 & 1.10667 & 3.59999 \\
 (b)  & 2         & 3         &  2         & 2        & 1.0697           & ---    & ---      & 1.03519 & ---     \\
 (c)  & 1         & 4.5       &  2         & 1.99     & 1.71             & 9.13   & 0.956    & 1.67    & 10.7    \\
 \hline
 \end{tabular}
  \caption{Comparison of 1VEV and 2VEV vortices close to the boundary: (a) $e_1=e_2=1$, (b) $e_1=1$, $e_2=0$, and (c) global. The coefficient of the $2\pi\log(R/R_c)$ term in the energy, $E_L$ is always 1 for 1VEV global vortices, and it is displayed for the $e_1=e_2=1$ case in the table, where $R_{\rm core}$ is defined as $a(R_{\rm core})=0.95 a(r\to\infty)$, for the global case, $f_1(R_{\rm core})=0.95 f_1(r\to\infty)$.}
  \label{tab:trans}
\end{table}

\begin{table}
  \centering
  \begin{tabular}{|c|c|c|c||c|}
   \hline
   $\beta_1$  & $\beta_2$ & $\beta'$ & $\alpha$  & $E/(2\pi)$\\
   \hline\hline
   2          & 3         & 2        & 2.011     & 1.03519 \\
   2          & 3         & 2        & 2.1       & 0.90588 \\
   2          & 12        & 4        & 4.2       & 0.87461 \\
   2          & 27        & 6        & 6.3       & 0.86128 \\
   2          & 48        & 8        & 8.4       & 0.85389 \\
   \hline
  \end{tabular}
 \caption{The energy of some 2VEV vortices for $e_2=0$, $n=1$, $m=0$.}
 \label{tab:2VEVe20}
\end{table}

\begin{table}
  \centering
  \begin{tabular}{|c|c|c|c||c|}
   \hline
   $\beta_1$  & $\beta_2$ & $\beta'$ & $\alpha$  & $E/(4\pi)$\\
   \hline\hline
   2          & 3         & 2        & 2.011     & 0.98606 \\
   2          & 3         & 2        & 2.1       & 0.86958 \\
   2          & 12        & 4        & 4.2       & 0.84219 \\
   2          & 27        & 6        & 6.3       & 0.83159 \\
   2          & 48        & 8        & 8.4       & 0.82593 \\
   \hline
  \end{tabular}
 \caption{The energy of some 2VEV vortices for $e_2=0$, $n=2$, $m=0$.}
 \label{tab:2VEVe202}
\end{table}

\section{Conclusions}\label{sec:conclusions}
In the present paper, we gave a detailed study of vortex solutions in a broad class of $U(1) \times U(1)$ symmetric, two-component scalar field theories. We emphasize the hitherto unexplored case, when one of the scalars obtains a vacuum expectation value (1VEV), and also consider the case with both fields having a VEV (2VEV).

In the 1VEV case of the purely scalar (Gross--Pitaevskii) theory, vortices can lower their energy by the formation of a condensate of the second field in their core. The result is a condensate core (CC) vortex. We found that the condensate in the core of the vortex can stabilise higher winding vortices against the splitting instability, in strong contrast with the
ordinary GP theory.

In the 1VEV case of the gauged theory, two-component Ginzburg--Landau theory (or in the relativistic case, extended Abelian Higgs model), CC vortices also exist. They coexist with embedded Abrikosov vortices, and have significantly lower energy. Importantly, CC vortices are stable. Higher flux CC vortices also stabilise against the splitting instability, even in such cases when embedded Abrikosov vortices split into unit flux ones. In a strong coupling limit, relevant to, e.g., superconducting liquid metallic hydrogen,, we have demonstrated the existence of stable ``giant'' vortices, i.e., vortices with $O(1000)$ flux quanta. The physical implication is that these materials are neither type II superconductors (which only have stable unit flux vortices), nor type I (as the energy/flux of vortices does have a minimum here).  We obtained similar results in the case when only one of the scalar fields is charged. In this case, we have found a remarkably simple description of the high flux limit of CC vortices,
quite similar to that of Abrikosov vortices.

In all the three cases of the GP and the GL models with 1 or 2 charged field, we have demonstrated, that vortices in the 1VEV case are smoothly connected with the ones in the 2VEV case with the winding in only one component. As in the case of two charged fields, the energy of 1VEV vortices is finite, and that of the corresponding 2VEV ones is logarithmically divergent, this connection is quite remarkable. In the case of one charged and one neutral field, all 1VEV and 2VEV vortices have finite energy.

The fact that CC vortices with higher fluxes become stable also implies a richer physics of inter-vortex forces. E.g., in the case of the GL theory, the stability of higher winding vortices implies that the inter-vortex forces become attractive as the distance between the vortices decreases. This is analogous to the behaviour of vortices in certain neither type I nor type II superconductors.

\subsection*{Acknowledgements}
This work has been supported by the grants OTKA K101709.

\appendix
\section{Details of the perturbation equations}\label{app:pertdetails}
We have studied the stability of the solutions at the linear level, writing the perturbed solution as $\phi_a=\phi_{a,\rm bg} +\epsilon \delta\phi_a$
and $A_\mu = A_{\mu,\rm bg} + \epsilon \delta A_\mu$. Here, the fields with index ``bg'' denote the static solution. In what follows, the ``bg'' index will be dropped.
Here we repeat the analysis of Ref.\ \cite{FL}, with the modified potential, and using the methods of Ref.\ \cite{Goodband}.
For the (in)stability analysis of the SU(2) symmetric case, see also Refs.\ \cite{twistedinstab1, FL, twistedinstab2}.

The linearised perturbation equations are cast into a form
\begin{equation}
  \label{eq:perteq}
  D \Psi = 0\,,\quad
\Psi =
\begin{pmatrix}
    \delta\phi_a \\ \delta\phi^*_a \\ \delta A^\mu
\end{pmatrix} = 0,
\end{equation}
where the operator $D$ is calculated from the linearised field equations. Simply linearising the field equations we would
get
\begin{equation}
  \label{eq:lineq}
D=
\begin{pmatrix}
D_{ab} - V_{a^*b}   & -V_{a^*b^*}         & B_{b\mu}   \\
-V_{ab}             & D_{ab}^* -V_{ab^*}  & B_{b\mu}^* \\
\tilde{B}_{a\mu}^*  & \tilde{B}_{a\mu}    & g_{\mu\nu}(\square +U_A) -\partial_\mu\partial_\nu
\end{pmatrix}\,,
\end{equation}
with (no implicit summations over indices $a$, $b$) $D_{ab}=\delta_{ab}(-\square + e_a^2 A^2) + 2ie_a\delta_{ab}A^\mu\partial_\mu$,
$\square=\partial_\mu \partial^\mu$, $V_{ab}=\partial^2V/\partial\phi_a\partial\phi_b$,  $V_{a^*b}=\partial^2V/\partial\phi_a^*\partial\phi_b$,
$V_{ab^*}=\partial^2V/\partial\phi_a\partial\phi_b^*$, $V_{a^*b^*} = \partial^2V/\partial\phi_a^*\partial\phi_b^*$, $U_A = 2\sum_a e_a^2 |\phi_a^2|$,
$B_{b\mu} = 2ie_b D_\mu \phi_b +ie_b\phi_b \partial_\mu$ and $\tilde{B}_{b\mu} = ie_b \partial_\mu \phi_b +2 e_b^2 A_\mu \phi_b -ie_b \phi_b \partial_\nu$.
However, the gauge condition can also be perturbed. In the background field gauge,
\begin{equation}
  \label{eq:BKGRGauge}
  F(A) := \partial_\mu \delta A^\mu + i\sum_a e_a (\delta\phi^*_a \phi_a - \phi^*_a\delta\phi_a) =0.
\end{equation}
which is added to the matrix in Eq.\ (\ref{eq:lineq}), to each line as $i e_a \phi_a F$, $-ie_a \phi_a^* F$, $\partial_\mu F$ (no summation over $a$), yielding
\begin{equation}
  \label{eq:perteq2}
D'=
\begin{pmatrix}
D_{ab}' -V_{a^*b}                    & - V_{a^*b^*} + e_a \phi_a e_b \phi_b    &  2i D_\mu \phi_b \\
-V_{ab} + e_a \phi_a^* e_b \phi_b^*  & D_{ab}^*{}' - V_{ab^*}                  & -2i(D_\mu\phi_b)^* \\
-2i(D_\nu\phi_a)^*                   & 2i D_\nu\phi_a                          & g_{\mu\nu}(\square+U_A)
\end{pmatrix}\,,
\end{equation}
where $D_{ab}'=D_{ab} - e_a\phi_ae_b\phi_b^*$.

The temporal component of the gauge field, $\delta A_0$ satisfies a decoupled equation
\begin{equation}
  \label{eq:FRVa0}
  (\square + U_A) \delta A_0 = 0\,,
\end{equation}
which agrees with the equation satisfied by the generators of infinitesimal gauge transformations,
\begin{equation}
  \label{eq:infgauge}
  \begin{aligned}
  \delta \phi_a &\rightarrow \delta \phi_a + i e_a\chi\phi_a,\\
  \delta A_\mu  &\rightarrow \delta A_\mu + \partial_\mu \chi,
  \end{aligned}
\end{equation}
that are still allowed by the gauge fixing (\ref{eq:BKGRGauge}), the ghost modes,
\begin{equation}\label{eq:FRVghost}
(\square+U_A)\chi =0\,.
\end{equation}
The ghost modes cancel the $\delta A_0$ spectrum, and a part of the spectrum of the remaining components of Eq.\ (\ref{eq:perteq}).

Equation (\ref{eq:perteq}) can be brought to the form of an eigenvalue equation by Fourier transforming in the $t$ and $z$ variables. For more
details, see Appendix \ref{app:pertdetails}. The resulting equations take the form
\begin{equation}
  \label{eq:perteq-fourier}
  \mathcal{M}  \tilde\Psi = \Omega^2 \tilde\Psi.
\end{equation}
where the $z$ direction wave number, $k$ is a parameter in $\mathcal{M}$. By expanding the fields in Fourier components in the angular variable, $\vartheta$,
separate eigenvalue equations
\begin{equation}
  \label{eq:FRVell}
  M_\ell  \Psi_\ell = \Omega^2 \Psi_\ell\,,
\end{equation}
are obtained for each partial wave $\ell$, where $M_\ell$ is an ordinary differential operator in the radial variable $r$. Similar treatment is possible for
the $\delta A_0$ and ghost modes. Here, $\Omega^2 < 0$ is the sign of an instability.

Equations (\ref{eq:FRVell}) possess a symmetry: replacing $k\to -k$, and exchanging the positive and negative frequency field components. This makes it possible to
examine only the $k>0$ region. We have solved Eq.\ (\ref{eq:FRVell}) with a slightly modified version of the shooting to a fitting point method of Ref.\ \cite{NR}.

We take one Fourier mode for the perturbations as
\begin{equation}
  \label{eq:FRV-fouriermodes}
  \begin{aligned}
    \delta \phi_1 (z,t;x_i) &= e^{i(\Omega t-k z)} \delta \phi_1 (k,\Omega;x_i) \\
    \delta \phi_2 (z,t;x_i) &= e^{i(\Omega t-(k-\omega) z)} \delta \phi_2 (k,\Omega;x_i) \\
    \delta A_\mu (z,t;x_i) &= e^{i(\Omega t-k z)} \delta A_\mu (k,\Omega;x_i)
  \end{aligned}
\qquad
  \begin{aligned}
    \delta \phi_1^* (z,t;x_i) &= e^{i(\Omega t-k z)} \delta \phi_1^* (-k,-\Omega;x_i) \\
    \delta \phi_2^* (z,t;x_i) &= e^{i(\Omega t-(k+\omega) z)} \delta \phi_2^* (-k,-\Omega;x_i) \\
    {}\\
  \end{aligned}
\end{equation}
with the index $i$ running over 1,2.
The variables $A_\mu$ are real functions, therefore
\begin{equation}
  A_\mu(k,\Omega,x_i) = A_\mu^*(-k,-\Omega,x_i).
\end{equation}
Substituting these into equations (\ref{eq:perteq}) yields the perturbation operator $\mathcal{M}$ of equation (\ref{eq:perteq-fourier}):
\begin{equation}
  \label{eq:fourM}
  \mathcal{M}=\begin{pmatrix}
    \mathcal{D}_1      & \mathcal{U}_1    & \mathcal{V}_1        & \mathcal{V}_1'   & \mathcal{A}_{1k}           & \mathcal{B}_1   & 0 \\
    \mathcal{U}_1^*    & \mathcal{D}_1^*  & \mathcal{V}_1'{}^{*} & \mathcal{V}_1^*  & \mathcal{A}_{1k}^*         & \mathcal{B}_1^* & 0 \\
    \mathcal{V}_2      & \mathcal{V}_2'   & \mathcal{D}_2        & \mathcal{U}_2    & \mathcal{A}_{2k}           & \mathcal{B}_2   & 0 \\
    \mathcal{V}_2'{}^* & \mathcal{V}_2^*  & \mathcal{U}_2^*      & \mathcal{D}_2^*  & \mathcal{A}_{2k}^*         & \mathcal{B}_2^* & 0 \\
    \mathcal{A}_{1i}^* & \mathcal{A}_{1i} & \mathcal{A}_{2i}^*   & \mathcal{A}_{2i} & \mathcal{D}_3\delta_{ik}   & 0               & 0 \\
    \mathcal{B}_{1i}^* & \mathcal{B}_{1i} & \mathcal{B}_{2i}^*   & \mathcal{B}_{2i} & 0                          & \mathcal{D}_3   & 0 \\
    0                  & 0                & 0                     & 0               & 0                          & 0               & \mathcal{D}_3 \\
  \end{pmatrix}
\end{equation}
with
\[\begin{aligned}
\mathcal{D}_1    &= k^2 -\partial_i^2 + e_1^2(A_i^2+A_3^2)
+ 2 i e_1 A_i \partial_i + 2 k e_1 A_3 +\mathcal{W}_1\,,\\
\mathcal{D}_1^*  &= k^2 
-\partial_i^2 + e_1^2(A_i^2+A_3^2)
- 2 i e_1 A_i \partial_i - 2 e_1 k A_3 +\mathcal{W}_1\,,\\
\mathcal{D}_2    &= (k-\omega)^2 
-\partial_i^2 + e_2^2(A_i^2+A_3^2)
+ 2i e_2 A_i \partial_i - 2 e_2 (\omega-k)A_3 +\mathcal{W}_2\,, \\
\mathcal{D}_2^*  &= (k+\omega)^2 
-\partial_i^2 + e_2^2(A_i^2+A_3^2)
- 2i e_2 A_i \partial_i - 2 e_2 (\omega+k)A_3 +\mathcal{W}_2\,, \\
\mathcal{D}_3   &= k^2 - \partial_i^2 
+ e_1^2 |\phi_1|^2 + e_2^2 |\phi_2|^2\,,
\end{aligned}\]
and (with no summation over $a$, $b$)
\[\begin{aligned}
\mathcal{W}_a\,  &= \frac{\partial^2 V}{\partial\phi_a^*\partial\phi_a} +e_a^2\phi_a^*\phi_a\\
\mathcal{U}_a\,  &= \frac{\partial^2 V}{\partial\phi_a^{*2}} -e_a^2\phi_a^2\\
\mathcal{V}_1\,  &= \frac{\partial^2V}{\partial \phi_1^*\phi_2} + e_1 e_2\phi_1\phi_2^*\\
\mathcal{V}_1'\, &= \frac{\partial^2V}{\partial \phi_1^*\phi_2^*} - e_1 e_2\phi_1\phi_2 \\
\mathcal{A}_{1i} &= 2 e_1 ( e_1 A_i \phi_1 + i\partial_i\phi_1 )\\
\mathcal{B}_{1i} &= 2 e_1^2 A_3 \phi_1
\end{aligned}\qquad
\begin{aligned}
\mathcal{W}^*_a\,  &= \frac{\partial^2 V}{\partial\phi_a^*\partial\phi_a} +e_a^2\phi_a^*\phi_a\\
\mathcal{U}^*_a\,  &= \frac{\partial^2 V}{\partial\phi_a^2} -e_a^2\phi_a^{*2}\\
\mathcal{V}_2\,    &= \frac{\partial^2V}{\partial \phi_2^*\phi_1}   - e_1 e_2\phi_2\phi_1^*\\
\mathcal{V}_2'\,   &= \frac{\partial^2V}{\partial \phi_2^*\phi_1^*} - e_1 e_2 \phi_2^*\phi_1^*\\
\mathcal{A}_{2i}   &= 2 e_2 ( e_2 A_i \phi_2 + i\partial_i\phi_2 )\\
\mathcal{B}_{2i}   &= 2 e_2 ( e_2 A_3-\omega) \phi_2
\end{aligned}\]
The Fourier-transform takes the equation (\ref{eq:FRVghost}) of the ghost modes into
\begin{equation}
  \label{eq:ghost-four}
  \mathcal{D}_3 \chi = \Omega^2 \chi,
\end{equation}
while a gauge transformation takes the form [see also Eq.\ (\ref{eq:infgauge})]
\begin{equation}
  \label{eq:gauge-four}
\delta A_i \to \delta A_i + \partial_i \chi,
\qquad \delta A_3 \to \delta A_3 - i k \chi, \qquad \delta A_0 \to \delta A_0 + i\Omega \chi.
\end{equation}
It is useful to introduce complex coordinates
\begin{equation}
  \label{eq:komplexkoord1}
  A_+ = \frac{e^{-i\vartheta}}{\sqrt{2}}(A_r - \frac{i}{r}A_\vartheta) \hspace{2.5cm} A_- = \frac{e^{i\vartheta}}{\sqrt{2}}(A_r + \frac{i}{r}A_\vartheta).
\end{equation}
Fourier expansion in the angle variable  (in cylindrical coordinates $x^1=r, x^2=\vartheta$, see Eq.\ (\ref{eq:perteq-fourier})),
omitting the sum over $\ell$ yields
\begin{equation}
  \label{eq:FRVpertans}
  \begin{aligned}
    \delta\phi_1 (k,\Omega)  &= s_{1,\ell} e^{i(n+\ell)\vartheta}\,, \\
    \delta\phi_2 (k,\Omega)  &= s_{2,\ell} e^{i(m+\ell)\vartheta}\,, \\
    \delta A_+   (k,\Omega)  &= i a_\ell e^{i(\ell-1)\vartheta}\,, \\
    \delta A_0   (k,\Omega)  &= a_{0,\ell} e^{i\ell\vartheta}\,,\\
    \delta A_3   (k,\Omega)  &= a_{3,\ell} e^{i\ell\vartheta}\,,\\
  \end{aligned}
  \quad\quad\quad
  \begin{aligned}
    \delta\phi_1^* (-k,-\Omega) &= s_{1,-\ell}^* e^{-i(n-l)\vartheta}\,, \\
    \delta\phi_2^* (-k,-\Omega) &= s_{2,-\ell}^* e^{-i(m-l)\vartheta}\,, \\
    \delta A_-   (-k,-\Omega)     &= -i a_{-\ell}^*  e^{i(\ell+1)\vartheta}\,, \\
    a_{0,-\ell}^* (-k,-\Omega)    &= a_{0,\ell}(k,\Omega)\,, \\
    a_{3,-\ell}^* (-k,-\Omega)    &= a_{3,\ell}(k,\Omega)\,. \\
  \end{aligned}
\end{equation}
Substituting this into the equations of motion, (\ref{eq:perteq}) assumes the form of an eigenvalue problem (\ref{eq:FRVell}) with
the operator
\begin{equation}
  M_\ell=
  \label{eq:FRVpertEig}
  \begin{pmatrix}
D_1  & U_1   & V    & V'    & A_1  & A_1'  & B_1  & 0  \\
U_1  & D_1^* & V'   & V     & A_1' & A_1   & B_1  & 0  \\
V    & V'    & D_2  & U_2   & A_2  & A_2'  & B_2  & 0  \\
V'   & V     & U_2  & D_2^* & A_2' & A_2   & B_2  & 0  \\
A_1  & A_1'  & A_2  & A_2'  & D_3  & 0     & 0    & 0  \\
A_1' & A_1   & A_2' & A_2   & 0    & D_3^* & 0    & 0  \\
B_1  & B_1   & B_2  & B_2   & 0    & 0     & D_4  & 0  \\
0    & 0     & 0   & 0     & 0    & 0     & 0    & D_4
  \end{pmatrix}
\end{equation}
with
\begin{equation}
  \begin{aligned}
    D_1    &= 
    - \nabla_r^2 + (k - e_1 \omega a_3)^2
    + \frac{(n(1- e_1 a)+\ell)^2}{r^2} 
    + W_1 \\
    D_1^*  &= 
    - \nabla_r^2 + (k - e_1 \omega a_2)^2
    + \frac{(n(1 - e_1 a)-\ell)^2}{r^2} 
    + W_1 \\
    D_2    &= 
    - \nabla_r^2 + (k - \omega + e_2 \omega a_3 )^2
    + \frac{(m-e_2 na+\ell)^2}{r^2} 
    + W_2 \\
    D_2^*  &= 
    - \nabla_r^2 + (k + \omega - e_2 \omega a_3 )^2
    + \frac{(m-e_2 na-\ell)^2}{r^2} 
    + W_2 \\
    D_3    &= D_a +\frac{(\ell-1)^2}{r^2} \\
    D_3^*  &= D_a +\frac{(\ell+1)^2}{r^2} \\
    D_4    &= D_a +\frac{\ell^2}{r^2} \\
\end{aligned}\end{equation}
with
\[
D_a = -\nabla_r^2+k^2 + 2(e_1^2 f_1^2 + e_2^2 f_2^2)\,,
\]
and
\[\begin{aligned}
  W_1   &= (2\beta_1+e_1^2) f_1^2-\beta_1 +\beta'f_2^2            \\
  U_1   &= (\beta_1-e_1^2) f_1^2                                  \\
  V     &= (\beta'+e_1 e_2) f_1 f_2                                \\
  A_1   &= -\sqrt{2}e_1\left(f_1'-\frac{n f_1}{r}(1-e_1 a)\right)  \\
  A_1'  &= \sqrt{2}e_1\left(f_1'+\frac{n f_1}{r}(1-e_1 a)\right)   \\
  B_1   &= 2e_1^2\omega a_3 f_1
\end{aligned}\hspace{4cm}\begin{aligned}
  W_2   &=  (2\beta_2+e_2^2)f_2^2-\alpha + \beta' f_1^2           \\
  U_2   &=  (\beta_2-e_2^2) f_2^2 \\
  V'    &=  (\beta'-e_1e_2) f_1 f_2 \\
  A_2   &= -\sqrt{2}e_2\left(f_2' -\frac{m-e_2na}{r}f_2\right) \\
  A_2'  &= \sqrt{2}e_2\left(f_2' +\frac{m-e_2na}{r}f_2\right) \\
  B_2   &= 2 e_2\omega (e_2 a_3-1) f_2 .
\end{aligned}
\]
The expansion of the gauge transformation generator function can be chosen as
\begin{equation}
  \label{eq:chi-four}
  \chi = \chi_\ell e^{i\ell\vartheta}.
\end{equation}
Using this expansion, the ghost mode equation (\ref{eq:ghost-four}) assumes the form
\begin{equation}
  \label{eq:ghost-ell}
  D_4 \chi_\ell = \Omega^2 \chi_\ell.
\end{equation}
Gauge transformations satisfying the above equation act on the fields as
\begin{equation}
  \label{eq:ell-gtrf}
  \begin{aligned}
    s_{a,\ell} & \to  s_{a,\ell} + i e_a \chi_\ell f_a\,, \\
    a_\ell    & \to  a_\ell - \frac{i}{\sqrt{2}}\left( \chi_\ell' +\frac{\ell\chi_\ell}{r} \right)\,, \\
    a_{3,\ell} & \to a_{3,\ell} - i k \chi_\ell\,,
  \end{aligned}
  \hspace{3.5cm}
  \begin{aligned}
    s_{a,-\ell}^* & \to  s_{a,-\ell}^* - i e_a \chi_\ell f_a\,, \\
    a_{-\ell}^*   & \to  a_{-\ell}^* + \frac{i}{\sqrt{2}}\left( \chi_\ell' -\frac{\ell\chi_\ell}{r} \right)\,, \\
    a_{0,\ell} & \to a_{3,\ell} + i \Omega \chi_\ell\,.
  \end{aligned}
\end{equation}

\section{Numerical data}\label{sec:numdata}
Some numerical data is presented here, in Tabs.\ \ref{tab:instab1}--\ref{tab:twb25}.

\begin{table}[h!]
\begin{center}
\begin{tabular}{|c|c|c|c|}
\hline
\multicolumn{4}{|c|}{$\beta_1=\beta_2=\alpha=2$, $\beta'=2.1$}\\
\hline
$\omega$ & $\Omega^2_{\rm min}$ & $\Omega_2^2$ & $k_{\rm min}$ \\
\hline\hline
ANO      & -0.03784           & 1            & 0.19453    \\
0.15     & $-9.7496\times 10^{-3}$ & 0.8954  & 0.1318     \\
0.12     & $-6.50\times 10^{-4}$   & 0.467   & 0.0693     \\
0.10     & $-3.15\times 10^{-6}$   & 0.241   & 0.00082    \\
0        & $1.7357\times 10^{-5}$  & 1       & 0          \\
\hline
\end{tabular}
\end{center}
\caption{Parameters of the dispersion relation (\ref{eq:disprel}) for some twisted vortices}
\label{tab:instab1}
\end{table}

\begin{table}[h!]
\begin{center}
\begin{tabular}{|c|c|c|c|}
\hline
\multicolumn{4}{|c|}{$\beta_1=\beta_2=\alpha=1.25$, $\beta'=1.255$}\\
\hline
$\omega$ & $\Omega^2_{\rm min}$ & $\Omega_2^2$ & $k_{\rm min}$ \\
\hline\hline
ANO      & $-1.47199\times 10^{-2}$ & 1      & $0.121325$ \\
0.11725  & $-1.2585\times 10^{-2}$  & 0.9988 & 0.11476    \\
0.06     & $-8.84\times 10^{-4}$    & 0.339  & 0.0315     \\
0.05     & $-8.83\times 10^{-4}$    & 0.105  & $4.3\times 10^{-4}$\\
0        & $-1.12\times 10^{-3}$    & 1      & 0\\
\hline
\end{tabular}
\end{center}
\caption{Parameters of the dispersion relation (\ref{eq:disprel}) for some twisted vortices}
\label{tab:instab2}
\end{table}

\begin{table}[h!]
\begin{center}
\begin{tabular}{|c|c|c|c|}
\hline
\multicolumn{4}{|c|}{$\beta_1=\beta_2=\alpha=2.5$, $\beta'=2.6$}\\
\hline
$\omega$ & $\Omega^2_{\rm min}$ & $\Omega_2^2$ & $k_{\rm min}$ \\
\hline\hline
ANO      & 0.11375           & 1            & 0.33726 \\
0.3      & $-5.8768\times 10^{-2}$ & 0.87785 & 0.2764  \\
0.2      & $-1.5306\times 10^{-3}$ & 0.4041  & 0.1058 \\
0        & $-1.8097\times 10^{-6}$   & 1       & 0\\
\hline
\end{tabular}
\end{center}
\caption{Parameters of the dispersion relation (\ref{eq:disprel}) for some twisted vortices}
\label{tab:instab3}
\end{table}

\begin{table}[h!]
\begin{center}
\begin{tabular}{|c|c|}
\hline
Parameters of the potential                    & $\Omega^2$ \\
\hline\hline
$\beta_1=\beta_2=\alpha=2$, $\beta'=2.1$       & 1.82239 \\
$\beta_1=\beta_2=\alpha=2.5$, $\beta'=2.6$     & 1.91046 \\
$\beta_1=\beta_2=\alpha=1.25$, $\beta'=1.255$  & 1.82486 \\
\hline
\end{tabular}
\end{center}
\caption{Eigenvalues of the ghost mode, $\omega=0$}
\label{tab:ghost}
\end{table}

\begin{table}[h!]
\begin{center}
\begin{tabular}{|c|c|}
\hline
$\beta_1$                    & $\Omega^2$ \\
\hline\hline
2                            & 1.76100 \\
2.5                          & 1.81814 \\
1.25                         & 1.62442 \\
\hline
\end{tabular}
\end{center}
\caption{Eigenvalues of the ghost mode, ANO vortex}
\label{tab:ghost2}
\end{table}

\begin{table}[h!]
\begin{center}
\begin{tabular}{|c|c|c|c|}
\hline
$\beta$  &      $a''(0)/2$   &       $f'(0)$        &       $E$ \\
\hline\hline
1        &    0.5            & 0.85318             & 6.28319  \\
1.25     &    0.53485        & 0.92418             & 6.58251  \\
2        &    0.61657        & 1.09935             & 7.26814  \\
2.5      &    0.65959        & 1.19677             & 7.62088  \\
\hline
\end{tabular}
\end{center}
\caption{ANO vortex data, $n=1$}
\label{tab:ANO}
\end{table}

\begin{table}[h!]
\begin{center}
\begin{tabular}{|c|c|c|c|}
\hline
$\beta$  &      $a''(0)/2$   &       $f''(0)/2$    &       $E$ \\
\hline\hline
1        &     0.25000       &   0.47229           &  12.56637  \\
1.25     &     0.27089       &   0.54508           &  13.35402  \\
2        &     0.32096       &   0.74438           &  15.20000  \\
2.5      &     0.34791       &   0.86724           &  16.17163  \\
\hline
\end{tabular}
\end{center}
\caption{ANO vortex data, $n=2$}
\label{tab:ANO2}
\end{table}

\begin{table}[h!]
\begin{center}
\begin{tabular}{|c|c|c|c|c|c|c|}
\hline
$\omega$ &      $a''(0)/2$   & $a_3(0)$      &  $f_1'(0)$   &  $f_2(0)$  &   $E$           & $\mathcal{I}$ \\
\hline\hline
0.11725  &  0.51959          & 0.03378       &  0.90743     & 0.17753   &   6.58162       & 0.19277       \\
0.06     &  0.35432          & 0.36881       &  0.71521     & 0.60258   &   6.52913       & 1.5920        \\
0.05     &  0.33484          & 0.40528       &  0.69093     & 0.63316   &   6.51540       & 1.5430        \\
0        &  0.28653          & ---           &  0.62872     & 0.70236   &   6.47450       & 0             \\
\hline
\end{tabular}
\end{center}
\caption{Twisted vortex data, $n=1$, $\beta_1=\beta_2=\alpha=1.25$, $\beta'=1.255$.}
\label{tab:twb125}
\end{table}

\begin{table}[h!]
\begin{center}
\begin{tabular}{|c|c|c|c|c|c|c|}
\hline
$\omega$ &      $a''(0)/2$   & $a_3(0)$      &  $f_1'(0)$   &  $f_2(0)$  &   $E$           & $\mathcal{I}$ \\
\hline\hline
0.15     & 0.57116           & 0.08031       & 1.04126     &  0.31237   &  7.26350        & 0.37172       \\
0.12     & 0.54780           & 0.12046       & 1.01109     &  0.38262   &  7.25713        & 0.46340       \\
0.10     & 0.53523           & 0.14179       & 0.99476     &  0.41508   &  7.25243        & 0.46448       \\
0        & 5.06806           & ---           & 0.95761     &  0.47935   &  7.23828        & 0             \\
\hline
\end{tabular}
\end{center}
\caption{Twisted vortex data, $n=1$, $\beta_1=\beta_2=\alpha=2$, $\beta'=2.1$.}
\label{tab:twb2}
\end{table}

\begin{table}[h!]
\begin{center}
\begin{tabular}{|c|c|c|c|c|c|c|}
\hline
$\omega$ &      $a''(0)/2$   & $a_3(0)$      &  $f_1'(0)$   &  $f_2(0)$  &   $E$           & $\mathcal{I}$ \\
\hline\hline
0.3      & 0.61039           & 0.07881      & 1.12946      & 0.32736    &  7.61484        & 0.63949       \\
0.2      & 0.51103           & 0.23013      & 0.99166      & 0.55480    &  7.55848        & 1.46249       \\
0        & 0.43019           & ---          & 0.87740      & 0.67434    &  7.45589        & 0             \\
\hline
\end{tabular}
\end{center}
\caption{Twisted vortex data, $n=1$, $\beta_1=\beta_2=\alpha=2.5$, $\beta'=2.5$.}
\label{tab:twb25}
\end{table}

\clearpage
\def\refttl#1{{\it ``#1''}, }%


\begin{thebibliography}{99}
\newcommand{\NPB}{\sl Nucl.\ Phys.\ \bf B\,}
\newcommand{\PLB}{\sl Phys.\ Lett.\ \bf B\,}
\newcommand{\PL}{\sl Phys.\ Lett.\ \bf}
\newcommand{\PRD}{\sl Phys.\ Rev.\ \bf D\,}
\newcommand{\PRB}{\sl Phys.\ Rev.\ \bf B\,}
\newcommand{\PRL}{\sl Phys.\ Rev.\ Lett.\ \bf}
\newcommand{\PRe}{\sl Phys.\ Rept.\ \bf}
\newcommand{\SJNP}{\sl Sov.\ J.\ Nucl.\ Phys.\ \bf}
\newcommand{\RPP}{\sl Rep.\ Prog.\ Phys.\ \bf}
\newcommand{\CMP}{\sl Commun.\ Math.\ Phys.\ \bf}
\newcommand{\ZPhys}{\sl Zeitschr.\ Phys.\ \bf}
\newcommand{\JHEP}{\sl JHEP\ \bf}


\bibitem{ANO} A.A.~Abrikosov, \refttl{The magnetic properties of superconducting alloys} {\sl J.\ Phys.\ Chem.\ Solids} {\bf 2}, 199-208 (1957);
H.B.~Nielsen and P.~Olesen, \refttl{Vortex-line models for dual strings} {\sl Nucl.\ Phys.}, {\bf B 61} (1973) 45.

\bibitem{Pismen} {L.M.~Pismen}, {\it Vortices in Nonlinear Fields}, Oxford University Press, Oxford, 1999.

\bibitem{SBP} B.~Svistunov, E.~Babaev, N.~Prokof'ev, {\sl Superfluid states of matter}, CRC Press, Boca Raton, 2015.

\bibitem{FS} K.~Fossheim and A.~Sudb\o, {\sl Superconductivity. Physics and Applications}, John Wiley \& Sons, Chichester, 2004.

\bibitem{VS} A.~Vilenkin and E.P.S.~Shellard, {\sl Cosmic strings and other topological defects}, Cambridge University Press, Cambridge, 1994.

\bibitem{kibble} M.B.~Hindmarsh and T.W.B.~Kibble, \refttl{Cosmic strings} {\sl Rep.\ Prog.\ Phys.} {\bf 58}, 477 (1995)
\arxiv{hep-ph/9411342}.

\bibitem{ZD} M.E.~Zhitomirsky and V.-H.~Dao, \refttl{Ginzburg-Landau theory of vortices in a multigap superconductor} {\PRB 69}, 054508 (2004) \arxiv{cond-mat/0309372 }.

\bibitem{Moshchalkov} V.V.~Moshchalkov, M.~Menghini, T.~Nishio, Q.H.~Chen, A.V.~Silhanek, V.H.~Dao, L.F.~Chibotaru, N.D.~Zhigadlo, and J.~Karpinski,
\refttl{Type-1.5 superconductors} {\PRL 102}, (2009) 117001 \arxiv[cond-mat]{0902.0997}.

\bibitem{Ashcroft68} N.W.~Ashcroft, \refttl{Metallic hydrogen: a high-temperature superconductor?} {\PRL} {\bf 21}, 1748 (1968).
\bibitem{Ashcroft91} K.~Moulopoulos and N.W.~Ashcroft, \refttl{Generalized Coulomb pairing in the condensed state} {\PRL} {\bf 66}, 2915 (1991).
\bibitem{Ashcroft98} K.~Johnson and N.W.~Ashcroft, \refttl{Proton pairing, electron pairing, and the phase structure of dense hydrogen}
{\sl J.\ Phys.: Condens.\ Matter} {\bf 10}, 11135 (1998).
\bibitem{Ashcroft99} K.~Moulopoulos and N.W.~Ashcroft, \refttl{Coulomb interactions and generalized pairing} {\sl Phys.\ Rev.} {\bf 59}, 12309 (1999).
\bibitem{Ashcroft2000} N.W.~Ashcroft, \refttl{The hydrogen liquids} {\sl J.\ Phys.: Condens.\ Matter} {\bf 12} (2000) A129-A137.
\bibitem{Ashcroft2005} N.W.~Ashcroft, \refttl{Metallic superfluids} {\sl Journal of Low Temperature Physics} {\bf 139}, 711-726 (2005).


\bibitem{ET} M.I.~Eremets and I.A.~Troyan, \refttl{Conductive dense hydrogen} {\sl Nature Materials} {\bf 10}, 927--931 (2011).
\bibitem{hardpressed} I.~Amato, \refttl{Metallic hydrogen: Hard pressed} {\sl Nature} {\bf 486}, 174–176(2012).

\bibitem{SSBS} E.~Sm\o{}rgrav, J.~Smiseth, E.~Babaev, and A.~Sudb\o{}, \refttl{Observation of a metallic superfluid in a numerical experiment}
{\sl Phys.\ Rev.\ Lett.} {\bf 95}, 135301 (2005) \arxiv{cond-mat/0508286}.

\bibitem{simul} J.~Chen, X.-Z.~Li, Q.~Zhang, M.I.J.~Probert, C.J.~Pickard, R.J.~Needs, A.~Michaelides adn E.~Wang,
\refttl{Quantum simulation of low-temperature metallic liquid hydrogen} {\sl Nature Communications} {\bf 4}, 2064 (2013) \arxiv[cond-mat]{1212.4554}.

\bibitem{Kasamatsu} K.~Kasamatsu, M.~Tsubota, and M.~Ueda, \refttl{Vortices in multicomponent Bode-Einstein condensates}
{\sl Int.\ J.\ Mod.\ Phys.} {\bf B19}, 1835-1904 (2005) \arxiv{cond-mat/0505546}.

\bibitem{Mason} P.~Mason, \refttl{Calculating the properties of a coreless vortices in a two-component condensate}
{\sl Phys.\ Rev.\ \bf A88}, 043608 (2013).

\bibitem{IvashinPoluektov} A.P.~Ivashin and Y.M.~Poluektonv, \refttl{Composite structure of vortices in two-component Bose-Einstein condensates}
{\sl Open Phys.\ \bf 2015,13}: 290-297 (2015) \arxiv[cond-mat]{1503.00147}.

\bibitem{KasamatsuEtoNitta} K.~Kasamatsu, M.~Eto, and M.~Nitta, \refttl{Short-range intervortex interaction and interacting dynamics of half-quantized vortices in two-component Bose-Einstein condensates} {\sl Phys.\ Rev.\ \bf A93}, 013615 (2016) \arxiv[cond-mat]{1510.00139}.

\bibitem{Catelani} G.~Catelani and E.A.~Yuzbashyan, \refttl{Coreless vorticity in multicomponent Bose and Fermi superfluids}
{\PRB 81}, 033629 (2010) \arxiv[cond-mat]{0909.3851}.

\bibitem{Jones} P.B.~Jones, \refttl{Type I and two-gap superconductivity in neutron star magnetism} {\sl Mon.\ Not.\ R.\ Astron.\ Soc.} {\bf 371}, 1327-1333 (2006) \arxiv{astro-ph/0606754}.

\bibitem{BabaevF} E.~Babaev, \refttl{Vortices with fractional flux in two-gap superconductors and in extended Faddeev model} {\PRL} {\bf 89}, 067001 (2002)
\arxiv{cond-mat/0111192}.
\bibitem{BS} E.~Babaev and M.~Speight, \refttl{Semi-Meissner state and neither type-I nor type-II superconductivity in multicomponent systems} {\PRB 72} (2005) 180502
\arxiv{cond-mat/0411681}.
\bibitem{BCSt15} E.~Babaev, J.~Carlström, and M.~Speight, \refttl{Type-1.5 superconducting state from an intrinsic proximity effect in two-band superconductors}
{\PRL} {\bf 105}, 067003 (2010) \arxiv[cond-mat]{0910.1607}.
\bibitem{BJS} E.~Babaev, J.~J\"aykk\"a, and M.~Speight, \refttl{Magnetic Field delocalization and Flux Inversion in Fractional Vortices in Two-Component Superconductors} {\PRL 103} (2009) 237002 \arxiv[cond-mat]{0903.3339}.
\bibitem{Zou} L.~Zou, \refttl{Non-Abrikosov vortices in liquid metallic hydrogen} {\sl Phys.~Lett.\ \bf A} {\bf 377},2182 (2013) \arxiv[cond-mat]{1307.1575}.

\bibitem{BSAdet} E.~Babaev, A.~Sudb\o, and N.W.~Ashcroft, \refttl{Observability of a projected new state of matter: a metallic superfluid}
{\PRL} {\bf 95}, 105301 (2005) \arxiv{cond-mat/0507605}.

\bibitem{BS1} M.~Silaev and E.~Babaev, \refttl{Microscopic theory of type-1.5 superconductivity in multiband systems} {\PRB 84}, 094515 (2011) \arxiv[cond-mat]{1102.5734}.
\bibitem{BS2} M.~Silaev and E.~Babaev, \refttl{Microscopic derivation of two-component Ginzburg-Landau model and conditions of its
applicability in two-band systems} {\PRB 85}, 134514 (2012) \arxiv[cond-mat]{1110.1593}.

\bibitem{CBS} J.~Carlström, E.~Babaev, and M.~Speight, \refttl{Type-1.5 superconductivity in multiband systems: effects of interband couplings}
{\PRB 83}, 174509 (2011) \arxiv[cond-mat]{1009.2196}.

\bibitem{CGB} J.~Carlström, J.~Garaud, and E.~Babaev, \refttl{Semi-Meissner state and nonpairwise intervortex interactions in type-1.5 superconductors}
{\PRB} {\bf 84}, 134515 (2011) \arxiv[cond-mat]{1101.4599}.

\bibitem{GB} J.~Garaud and E.~Babaev, \refttl{Vortex matter in $U(1)\times U(1)\times \mathbb{Z}_2$ phase-separated superconducting condensates}
{\PRB} {\bf 90}, 214524 (2014) \arxiv[cond-mat]{1410.2985}.



\bibitem{Witten} E.~Witten, \refttl{Superconducting string} {\NPB 249} (1985) 557--592.

\bibitem{Peter} P.~Peter, \refttl{Superconducting cosmic string: Equation of state for spacelike and timelike current in the neutral limit}
{\sl Phys.\ Rev.} {\bf D45}, 1091 (1992).

\bibitem{Peter2} P.~Peter, \refttl{Low-mass current-carrying cosmic strings}
{\sl Phys.\ Rev.} {\bf D46}, 3322 (1992).

\bibitem{vac-ach} T.~Vachaspati and A.~Ach\'ucarro, \refttl{Semilocal cosmic strings} {\PRD 44} (1991) 3067.
\bibitem{hin1} M.~Hindmarsh, \refttl{Existence and stability of semilocal strings} {\PRL 68} (1992) 1263
\bibitem{hin2} M.~Hindmarsh, \refttl{Semilocal topological defects} {\NPB 392} (1993) 461-492 \arxiv{hep-ph/9206229}.
%
\bibitem{semilocal} A.~Ach\'ucarro and T.~Vachaspati, \refttl{Semilocal and electroweak strings} {\PRe 327} (2000) 427 \arxiv{hep-ph/9904229}.

\bibitem{JPV1} M.~James, L.~Perivolaropoulos, and T.~Vachaspati,
\refttl{Stability of electroweak strings} {\sl Phys.\ Rev.} {\bf D46}, 5232 (1992).

\bibitem{JPV2} M.~James, L.~Perivolaropoulos, and T.~Vachaspati,
\refttl{Detailed stability analysis of electroweak strings} {\sl Nucl.\ Phys.} {\bf B395}, 534-546  (1993) \arxiv{hep-ph/9212301}.

\bibitem{Perkins} W.B.~Perkins, \refttl{$W$ condensation in electroweak strings}
{\sl Phys.\ Rev.} {\bf D47}, 5224  (1993).

\bibitem{GHelectroweak} M.~Goodband and M.~Hindmarsh, \refttl{Instabilities of electroweak strings}
{\sl Phys.\ Lett.} {\bf B363}, 58-64 (1995) \arxiv{hep-ph/9505357}.

\bibitem{FRV1} P.~Forgács, S.~Reuillon, and M.S.~Volkov, \refttl{Superconducting vortices in semilocal models}
{\PRL \bf 96}, 041601 (2006) \arxiv{hep-th/0507246}.

\bibitem{FRV2} P.~Forgács, S.~Reuillon, and M.S.~Volkov, \refttl{Twisted superconducting semilocal strings} {\NPB 751} (2006) 390--418 \arxiv{hep-th/0602175}.

\bibitem{GVelectroweak} J.~Garaud and M.S.~Volkov, \refttl{Superconducting non-Abelian vortices
in Weinberg--Salam theory --- electroweak thunderbolts}
{\NPB 826} (2010) 174--216, \arxiv[hep-th]{0906.2996}.

\bibitem{twistedinstab1} J.~Garaud and M.S.~Volkov, \refttl{Stability analysis of the twisted superconducting semilocal strings}
{\sl Nucl.\ Phys.} {\bf B799} (2008) 430-455 \arxiv[hep-th]{0712.3589}

\bibitem{FL} P.~Forgács, Á.~Lukács, \refttl{Instabilities of twisted strings} {\sl JHEP} {\bf 0912} (2009) 064 \arxiv[hep-th]{0908.2621}.

\bibitem{twistedinstab2} B.~Hartmann, P.~Peter, \refttl{Can type II Semi-local cosmic strings form?} {\sl  Phys.\ Rev.} {\bf D86} (2012) 103516  \arxiv[hep-th]{1204.1270}

\bibitem{GVinstab} J.~Garaud and M.S.~Volkov, \refttl{Stability analysis of superconducting electroweak vortices}
{\NPB 839} (2010) 310--340, \arxiv[hep-th]{1005.3002}.

\bibitem{Erice} Á.~Lukács, \refttl{Twisted strings in Extended Abelian Higgs Models} In: A.~Zichichi (ed): {\sl What is known and unexpected at the LHC},
Proceedings of the International School of Subnuclear Physics, Erice-Sicily, Italy, 29 August -- 7 September 2010, World Scientific, 2013.

\bibitem{FL2} P.~Forgács and Á.~Lukács, \refttl{Vortices with scalar condensates in two-component Ginzburg-Landau systems} \arxiv[hep-th]{1603.03291}.


\bibitem{SilveiraZee} V.~Silveira, A.~Zee, \refttl{Scalar phantoms} {\sl Phys.\ Lett.} {\bf 161B}, 136-140 (1985).

\bibitem{PW} B.~Patt, F.~Wilczek, \refttl{Higgs-field portal into hidden sectors} \arxiv{hep-ph/0605188}.

\bibitem{VachaspatiWatkins} T.~Vachaspati and R.~Watkins, \refttl{Bound states can stabilize electroweak strings}
{\sl Phys.\ Lett.} {\bf B318}, 163-168 (1993) \arxiv{hep-ph/9211284}.

\bibitem{PerivolaropoulosPlatis} L.~Perivolaropoulos and N.~Platis,
\refttl{Stabilizing the semilocal string with a dilatonic coupling} {\PRD 88}, 065017 (2013) \arxiv[hep-ph]{1307.3920}.

\bibitem{HartmannArbabzadah} B.~Hartmann and F.~Arbabzadah, \refttl{Cosmic strings interacting with dark strings}
{\sl JHEP} {\bf 07} (2009) 068 \arxiv[hep-th]{0904.4591}.

\bibitem{BrihayeHartmann} Y.~Brihaye and B.~Hartmann, \refttl{Effect of dark strings on semilocal strings}
{\sl Phys.\ Rev.} {\bf  D 80}, 123502 (2009) \arxiv[hep-th]{0907.3233}.


\bibitem{StringsSprings} D.~Haws, M.~Hindmarsh, and N.~Turok, \refttl{Superconducting strings or springs?}
{\sl Phys.\ Lett.\ \bf B209}, 255-261 (1988).

\bibitem{Landau3} L.D.~Landau and E.M.~Lifshitz, {\it Course of Theoretical Physics 3.\ Quantum Mechanics}, Pergamon Press, Oxford, 1977.

\bibitem{Goodband} M.~Goodband and M.~Hindmarsh, \refttl{Bound states and instabilities of vortices}
{Phys.\ Rev.} {\bf D52} 4621 \arxiv{hep-ph/9503457}.

\bibitem{FM} P.~Forg\'acs and N.S.~Manton, \refttl{Space-time symmetries in gauge theories} {\CMP 72} 15 (1980).

\bibitem{NR} W.H.~Press, S.A.~Teukolsky, W.T.~Vetterling, and B.P.~Flannery,
{\sl Numerical Recipes 3rd Edition: The Art of Scientific Computing}, Cambridge University Press, Cambridge, 2007.

\bibitem{Bolognesi1} S.~Bolognesi, \refttl{Domain walls and flux tubes} {\sl Nucl.\ Phys.} {\bf B730} (2005) 127-149 \arxiv{hep-th/0507273}.

\bibitem{Bolognesi2} S.~Bolognesi and S.B.~Gudnason, \refttl{Multi-vortices are wall vortices: A numerical proof} {\sl Nucl.\ Phys} {\bf B741} (2006) 1-16
\arxiv{hep-th/0512132}.
\end{thebibliography}
\end{document}